\documentclass[%
 reprint,
 amsmath,amssymb,
 aps,
]{revtex4-1}

\usepackage{graphicx}
\usepackage{appendix}
\usepackage{varioref}
\usepackage{cleveref}
\usepackage{subfig}
\usepackage{dcolumn}
\usepackage{bm}
\usepackage[mathlines]{lineno}

\usepackage{xcolor}

\begin{document}


\title{Monopoles, vortices and their correlations in SU($3$) gauge group}

\author{N. Karimimanesh}
 \altaffiliation{karimi.n@ut.ac.ir}
\author{S. Deldar}%
 \email{sdeldar@ut.ac.ir}
 \author{Z. Asmaee}%
 \email{zahra.asmaee@ut.ac.ir}
\affiliation{Department of Physics, University of Tehran,\\
P.O. Box 14395-547, Tehran 1439955961, Iran}

\begin{abstract}

Topological defects such as monopoles, vortices and \textquotedblleft chains\textquotedblright of the SU($3$) gauge group are studied using its SU($2$) subgroups. 
Two appropriate successive gauge 
transformations are applied to the subgroups to identify the chains of monopoles and vortices. Using the fact that the defects of the subgroups are not 
independent, the SU($3$) defects and the Lagrangian are studied and compared with the Cho decomposition method obtained for monopoles.  
By comparing the results with the ones which are obtained directly for the SU($3$) gauge group, 
the relations and the possible interactions between the defects of the subgroups are discussed.

\begin{description}
\item[PACS numbers]
14.80.Hv, 12.38.Aw, 12.38.Lg, 11.15.-q
\end{description}
\end{abstract}

\pacs{Valid PACS appear here}
\maketitle


\section{\label{sec:level1}INTRODUCTION}

Since the time of discovering quarks, there have been so many attempts to describe the confinement problem but yet no universal, comprehensive and analytical description 
based on the basic principles of quantum Chromodynamics to explain some more expected features of the confinement. Lattice calculations have given 
valuable clues to understand this phenomenon and to help proposing appropriate phenomenological models. Monopoles and vortices are among the rather successful 
candidates in describing the confinement problem, both via lattice calculations and phenomenological models. However, none of these candidates have been able to interpret all the characteristics of confinement like the Casimir scaling and N-altiy dependence. On the other hand, lattice results have shown some indications of field
configurations of correlated vortices and monopoles, where center vortices have been observed to end at monopole world-lines.
In fact, numerical simulations in the indirect
maximal center gauge have found that monopole world lines lie
on center vortex sheets and
the magnetic flux of those monopole is concentrated along the sheet.
In other words, the magnetic flux of monopoles at given time is
concentrated in a tube-like structure, and the monopole \textquotedblleft chain\textquotedblright\  is
shorthand terminology for an ordering of monopoles and
antimonopoles, in which half the magnetic flux from a monopole ends on
the next antimonopole in the chain, while the other half ends on the
previous antimonopole. 
 Some  
phenomenological models have been presented to include both defects in the form of chains of monopoles and vortices. People hope that chains of monopoles and vortices 
may describe some more expected characteristics of the quark-antiquark potential in the confinement regime. Some of the papers in this regard are given in 
references \cite{pepe} to \cite{oxman19}. We have also studied chains of monopoles and vortices and the corresponding field strength tensors and the Lagrangian 
for the SU($2$) gauge group in ref. \cite{karimi}.
In this article, we generalize our previous article to contain the SU($3$) gauge group and to study the possible defects and their interactions.

Basically, in many of the phenomenological models 
which try to describe quark confinement, one looks for magnetic topological objects or defects in the vacuum of QCD. There are a couple of ways to identify 
these magnetic defects. One of them is the field decomposition method proposed by Duan, Ge, Cho, Faddeev, Niemi and shabanov \cite{duan, cho1, cho2, cho3, fadev, shaban} 
where the Yang-Mills field is decomposed into other appropriate variables. As a result, topological defects such as monopoles appear in the theory. 
What is noticeable about the Cho decomposition method, is the fact that it allows
a straightforward generalization of the SU($2$) results to the SU($N$) in a Weyl symmetric form. In refs. \cite{massgap, 2019} using the Weyl symmetry, 
Abelian (Cho-Duan-Ge) 
decomposition of the SU($3$) gauge field is discussed.  
Abelian gauge fixing is another method in which the formation of monopoles in the QCD ground state for SU($2$) and SU($3$) gauge groups are 
discussed \cite{Ripka}.

The other clever idea to identify magnetic defects, is the use of appropriate gauge transformations to transform the color frames in such a way that they reveal monopoles, 
vortices, or both. Using this method, monopoles, vortices, and chains for the SU($2$) gauge group are introduced \cite{oxman, karimi}. 
In this article, given the fact that the SU($3$) group contains three SU($2$) subgroups, we generalize this method to the SU($3$) gauge group and 
identify monopoles, vortices and their correlations.
Using SU($2$) subgroups makes the task much easier than using directly the SU($3$) gauge group. In addition, identifying SU($3$) defects from the SU($2$) 
subgroups, the relations between topological defects of the three SU($2$) subgroups are discussed.

Here, we present an overview of our goals and the steps we go through.
Using the appropriate gauge transformations for each of the three SU($2$) subgroups, we obtain the local frames which contain defects such as monopoles, vortices or chains.  
Applying these gauge transformations, the transformed gauge fields look like the Cho decomposition fields  for monopoles. In fact, we are dealing with Cho decomposition written based on local frames containing defects. 
We indicate the relationship between the magnetic defects obtained from the gauge transformations 
on the SU($2$) subgroups and the results obtained by Cho decomposition, using directly the SU($3$) gauge group.
Then, we write the Lagrangian in two forms:
in terms of the three SU($2$) subgroups where we can learn 
how topological objects of each of the three SU($2$) subgroups interact with each other;
and by constructing the SU($3$) Lagrangian from its SU($2$) subgroups to study the interaction between topological objects in SU($3$) group.
 These information may be used to write an effective Lagrangian to obtain the confinement potential using topological defects.

In Section \ref{sec:level2} we review the above gauge transformation method for identifying monopoles, vortices and chains for the SU($2$) gauge group. 
In Section \ref{sec:level4} we first apply the appropriate gauge transformations to introduce the monopoles of each of the SU($2$) subgroups. 
Then, using the results of each subgroup, we introduce the new local color frames and discuss the connection between the applied method and the Cho decomposition method.
In addition, using the results of the Cho decomposition and the fact that the monopoles of the three SU($2$) subgroups are not independent, we study
the relations between the monopoles of three SU($2$) subgroups and the monopoles of SU($3$) group.
Similar to the general procedure of Section \ref{sec:level4}, in Section \ref{sec:level5} we identify the vortices for SU($3$) gauge group.
In Section \ref{sec:level6}, successively applying the two gauge transformations introduced in Sections \ref{sec:level4} and \ref{sec:level5}, we argue 
about the correlation between monopoles and vortices of the three SU($2$) subgroups.
We obtain local frames containing both monopole and vortex. The transformed gauge field is very similar to the Cho decomposition gauge fields  for monopoles. 
Therefore, the results of the Cho decomposition can be used for finding out the relation between the correlated monopoles and vortices of the three SU($2$) 
subgroups and the correlated monopoles and vortices of SU($3$) group. 
At the end, we study the Lagrangian and discuss about the possible interactions between the defects. 
The conclusion and summary are given in Section \ref{sec:level7}.

\section{\label{sec:level2}MONOPOLES AND CENTER VORTICES AS DEFECTS OF THE LOCAL COLOR FRAMES}

By appropriate gauge transformations, we obtain the local color frames that contain the possible defects of the theory. As a result, the gluon fields are decomposed 
in such a way that the contribution of the magnetic defects appears explicitly in the final gauge field.
 
\subsection{Monopole}

In contrast to Dirac's motivation that added monopoles to the Maxwell's equation from the point of view of aesthetic, monopoles may appear in QCD
as the effective agents to describe the confinement problem. Even though they do not exist in the standard form of the Lagrangian, it is possible to
identify them by various methods like field decompositions and projections. Their existence has been confirmed by lattice calculations, as well.
For example, as a result of the decomposition, one expects to observe the Wu-Yang monopoles which are point-like topological objects. 
There are some different ways to identify these defects. 
One of these methods is to use a nontrivial gauge transformation to construct the local color frame that contains monopoles. 

By applying a nontrivial gauge transformation, the SU($2$) Yang-Mills gauge field and the field strength tensor are changed as the following,
\begin{equation}
\vec{A}_\mu ^U.\vec{T}=U\left( \vec{A}_\mu.\vec{T}\right) U^{-1}+\frac{i}{g}U\partial_\mu U^{-1},\label{a}
\end{equation}
\begin{equation}
\vec{F}_{\mu\nu}^U.\vec{T}=U\left( \vec{F}_{\mu\nu}.\vec{T}\right) U^{-1}+\frac{i}{g}U[\partial_\mu,\partial_\nu]U^{-1},\label{gauge}
\end{equation}
where $U\in SU(2)$ is a nontrivial topological mapping that is single-valued along any closed loop and the components of $\vec{T}$ are the generators of the SU($2$) group.

To make a connection with the Cho decomposition, we introduce the frame $\hat{m}_a, a=1,2,3$, which is constructed by applying a rotation $R(U)$ on the 
basis $\hat{e}_a$ of the color space,
\begin{equation}
UT_a U^{-1}=\hat{m}_a.\vec{T}\   \   \text{or},   \   \hat{m}_a=R(U)\hat{e}_a.\label{ma}
\end{equation}
This gauge transformation can be expressed in terms of Euler angles,
\begin{equation}
U=e^{-i\alpha T_3}e^{-i\beta T_2}e^{-i\gamma T_3}\   \   ,   \   \ R(U)=e^{\alpha M_3}e^{\beta M_2}e^{\gamma M_3},   \label{39}
\end{equation}
where $M_a$'s are the generators of SO($3$) gauge group and $T_a$'s indicate half of the Pauli matrices.

In order for the frame $(\hat{m}_1,\hat{m}_2,\hat{m})$ to contain monopoles, we choose the parameters $\alpha$, $\beta$ and $\gamma$ such that 
$\alpha=\gamma=\varphi, \beta=\theta$, where $\theta$ and $\varphi$ are the polar and azimuthal angles. With this choice, the Abelian direction $\hat{m}$ takes the form of a 
hedgehog ($\hat{m}=\hat{r}$). The matrix form of the transformation $U$ and the explicit form of the vectors $\hat{m}_a$ are calculated in \cite{karimi}. 

Using the explicit forms of $T_a$'s in  $U$, the gauge field of Eqn. \eqref{a} is transformed as the following \cite{karimi, oxman},
\begin{equation}
\vec{A}_\mu^U.\vec{T}=[\underbrace{(A_\mu^3-C_\mu^{(m)})}_{=A_\mu^{(m)}}\hat{m}-\frac{1}{g}\hat{m}\times\partial_\mu \hat{m}+\underbrace{A_\mu^1\hat{m}_1+A_\mu^2\hat{m}_2}_{=\vec{X}_\mu^{(m)}}].\vec{T}, \label{A-vector}
\end{equation}
where,
\begin{equation}
C_\mu^{(m)}=-\frac{1}{g}\hat{m}_1.\partial_\mu\hat{m}_2.\label{Cmu}
\end{equation}
Using the gauge transformation $U$ with the chosen parameters given after Eqn. \eqref{39}, the local color direction $\hat{m}$ would be equal to $\hat{r}$.
Thus, choosing $A_\mu^{(m)}=0$, $A_\mu^a=0$ for $a=1,2$, a static Wu-Yang monopole \cite{wu} with a hedgehog form is obtained by Eqn. \eqref{A-vector}.

We would like to compare the gauge field $\vec{A}_\mu^U$ given by Eqn. (\ref{A-vector}) with the gauge field obtained from the Cho decomposition. It is 
very interesting that 
the form of the resulted gauge fields of these two methods are very similar to each other. In Cho decomposition, by imposing the magnetic isometry to the gauge 
potential $\vec{A}_\mu$, the restricted potential is obtained,
\begin{equation}
D_\mu\hat{m}=(\partial_\mu +g \vec{A}_\mu\times)\hat{m}=0\Rightarrow \hat{A}_\mu=A_\mu^{(m)}\hat{m}-\frac{1}{g}\hat{m}\times\partial_\mu\hat{m}.
\end{equation}
Choosing $A_\mu^{(m)}=0$ and $\hat{m}=\hat{r}$, the restricted potential would describe the Wu-Yang monopole. By 
adding $\vec{X}_\mu^{(m)}=A_\mu^1\hat{m}_1+A_\mu^2\hat{m}_2$ to the restricted potential, the extended Cho decomposition similar to what we have gotten in 
Eqn. \eqref{A-vector}, is obtained. The major difference between the two methods is that in the Cho decomposition method the direction $\hat{m}$ is chosen to be $\hat{r}$
after the decomposition is done, while in the method used in this article, we use a specific nontrivial gauge transformation and as a result the form of a hedgehog 
for $\hat{m}$ is obtained, at the end.

\subsection{vortex}

Center vortices are color magnetic defects which are localized on the closed two-dimensional surfaces (closed one-dimensional strings) in 4D (3D). Center 
vortices in SU($N$) Yang-Mills theory are defined by the center of the gauge group. If a center vortex is linked to a Wilson loop, the Wilson loop  
is multiplied by an element proportional to the center of Z($N$) group. Lattice calculations confirm that vortices are responsible for the color confinement \cite{mack} 
and removing them from the theory removes the linear behavior of the quark-antiquark potential.
\\

In the continuum, thin center vortices are introduced by applying a gauge transformation $V$\cite{engel,rein,thinv},
\begin{equation}
\vec{A}_\mu^V\cdot\vec{T}=VA_\mu^a T^aV^{-1}+\frac{i}{g}V\partial_\mu V^{-1}-I_\mu(\vartheta).\label{vartheta}
\end{equation}
For simplicity we study the SU($2$) group and choose $V\in SU(2)$. $I_\mu(\vartheta)$ represents the ideal vortex, and  is localized on three-volume $\vartheta$ 
whose border gives the closed thin center vortex worldsheet. When crossing a three-volume $\vartheta$, the mapping $V$ changes by a center element, $e^{\pm i\pi}V$. 
$I_\mu(\vartheta)$ is designed to cancel derivatives of the discontinuity of $V$ at $\vartheta$ and keeping only 
the effect of the border of $\vartheta$, called the thin center vortices.
\\

Using the gauge transformation $V$, a local basis $\hat{n}'_a$ in color space is introduced,
\begin{equation}
VT^aV^{-1}=\hat{n}'_a.\vec{T}\   \   ,   \   \hat{n}'_a=R_{\hat{n}'}\hat{e}_a,\label{S-Gauge}
\end{equation}
and the frame dependent fields,
\begin{equation}
C_\mu^{(v)a}=-\frac{1}{2g}\epsilon^{abc}\hat{n}'_b\cdot\partial_\mu\hat{n}'_c,
\end{equation}
are defined so that they satisfy the following properties,
\begin{equation} 
\hat{n}'_b\cdot\partial_\mu\hat{n}'_c=-g\epsilon^{abc}C_\mu^a\quad,\quad C_\mu^aM^a=\frac{i}{g}R^{-1}\partial_\mu R.\label{RR}
\end{equation}
The matrices $M^a$, with elements $(M^a)^{bc}=-i\epsilon^{abc}$ are the adjoint generators and $[M^a,M^b]=i\epsilon^{abc}M^c$, $Tr(M^aM^b)=2\delta^{ab}$. 
$\epsilon^{abc}$ is the Levi-Civita symbol. For SU($2$) gauge group, we have $V=e^{i\varphi T_3}$ or in adjoint representation $R_{\hat{n}'}=R_3(\varphi)$. 
Using the matrix form of  $V$ or its equivalent counterpart in the adjoint representation $R_3$, the components of $\hat{n}'_a,\  a=1,2,3$ are calculated\cite{karimi},
\begin{equation}
\hat{n}'_1=\left(\begin{array}{c}
\cos\varphi\\
-\sin\varphi\\

0
\end{array}\right),\quad\hat{n}'_2=\left(\begin{array}{c}
\sin\varphi\\
\cos\varphi\\
0
\end{array}\right),\quad\hat{n}'_3=\left(\begin{array}{c}
0\\
0\\
1
\end{array}\right).
\end{equation}
Using the components of $\hat{n}'_a$, the frame dependent fields are obtained,
\begin{equation}\begin{split}
C_\mu^{(v)1}&=-\frac{1}{g}\hat{n}'_2\cdot\partial_\mu\hat{n}'_3=0,\\
C_\mu^{(v)2}&=-\frac{1}{g}\hat{n}'_3\cdot\partial_\mu\hat{n}'_1=0,\\
C_\mu^{(v)3}&=-\frac{1}{g}\hat{n}'_1\cdot\partial_\mu\hat{n}'_2=-\frac{1}{g}\partial_\mu\varphi.\label{C-mu}
\end{split}\end{equation}

From the second equality of Eqn. \eqref{RR}, one can simply show that $-C_\mu^{(v)3}\hat{n}'_3\cdot\vec{T}=\frac{i}{g}V\partial_\mu V^{-1}$. However, 
unlike $V$, the adjoint representation $R$ and the corresponding frame $\hat{n}'_a$ are always continuous, and as a result $C_\mu^{(v)3}\hat{n}'_3$ contains no term 
concentrated on $\vartheta$\cite{thinv}. Therefore, one can rewrite the above equation as follows,
\begin{equation}
-C_\mu^{(v)3}\hat{n}'_3\cdot\vec{T}=\frac{i}{g}V\partial_\mu V^{-1}-I_\mu(\vartheta).\label{16}
\end{equation}
Replacing $\frac{i}{g}V\partial_\mu V^{-1}$ from Eqn. \eqref{16} in Eqn. \eqref{vartheta} and using the fact that $VA_\mu^aT_aV^{-1}=A_\mu^a\hat{n}'_a\cdot\vec{T}$, 
an equivalent representation for the thin configuration proposed in \cite{engel,rein} is obtained,
\begin{equation}
\vec{A}_\mu^V\cdot\vec{T}=\left[(A_\mu^3-C_\mu^{(v)3})\hat{n}'_3+A_\mu^1\hat{n}'_1+A_\mu^2\hat{n}'_2\right]\cdot\vec{T}.
\end{equation}
Using the definition of $C_\mu^{(v)3}$ of Eqn. \eqref{C-mu},
\begin{equation}
\vec{A}_\mu^V=(A_\mu^3+\frac{1}{g}\partial_\mu\varphi)\hat{n}'_3+A_\mu^1\hat{n}'_1+A_\mu^2\hat{n}'_2.
\end{equation}

\subsection{chain}

Since the scenarios written solely based on the monopoles or vortices have not been able to describe all the expected behaviors of the confining potential between a pair of 
quark and antiquark, it may be a smart idea to work with the configurations that include both of these objects. Chains are the configurations in which the vortex 
worldsheets have been attached to the monopole worldlines. Such configurations have been supported by lattice simulations \cite{greensite, pepe, zakharov}.

It has been shown that by implying two successive gauge transformations $U$ and $V$, one can observe a configuration that describes correlated monopoles and vortices or 
chains \cite{oxman},
\begin{equation}
(VU)T^a(VU)^{-1}=VUT^aU^{-1}V^{-1}=\hat{n}_a.\vec{T},
\end{equation}
\begin{equation}
\hat{n}_a=R(VU)\hat{e}_a=R(V)R(U)\hat{e}_a=R(V)\hat{m}_a.
\end{equation}
Using the above equations and the definition of Eqn. (\ref{Cmu}), $C_\mu=-\frac{1}{g}\hat{n}_1.\partial_\mu\hat{n}_2$ which is correct for any magnetic defect,
we have,
\begin{equation}
C_\mu^{(n)}=C_\mu^{(m)}+C_\mu^{(v)},\label{magnet}
\end{equation}
where,
\begin{equation}
C_\mu^{(m)}=-\frac{1}{g}\hat{m}_1.\partial_\mu\hat{m}_2,\   \    \    C_\mu^{(v)}=\frac{1}{g}\partial_\mu\varphi.
\end{equation}
The magnetic potential $C_\mu^{(n)}$ is a summation of two parts: $C_\mu^{(m)}$ which represents the magnetic potential of the monopole, and $C_\mu^{(v)}$ which represents the magnetic potential of the center vortex.

In \cite{karimi} we applied the gauge transformations introduced in this section to obtain monopole, vortex, and chain for SU($2$) gauge group and the 
corresponding Lagrangian. We have shown that 
the results obtained for monopoles are in agreement with what is obtained from Abelian Projection. We have also 
obtained a Lagrangian density for the correlated monopoles and vortices containing kinetic energy of the monopole, kinetic energy of the vortex, and the 
interaction between monopole and vortex. 

Using the fact that SU($3$) gauge group has three SU($2$) subgroups, we generalize the results obtained for the SU($2$) gauge group to the SU($3$) gauge group  
to study the monopoles, vortices and 
their correlations for this group. Using SU($2$) subgroups of SU($3$) group, people have found magnetic monopoles by both Abelian gauge fixing \cite{Ripka} and  
field decomposition method
\cite{massgap} for the SU($3$) gauge group. In this article, in addition to monopoles, we also study the vortex and the correlation between monopoles 
and vortices for the SU($3$) gauge group. In addition, by comparing the results obtained with the help of SU($2$) subgroups with the ones obtained directly from the 
Cho field decomposition for SU($3$) monopoles, we discuss about the connection between SU($2$) and SU($3$) defects. For this purpose, in the 
next section, we first introduce and categorize SU($2$) subgroups of SU($3$) gauge group and then, by applying appropriate gauge transformations, we construct local 
color frames for each of the subgroups. Depending on the type of gauge transformation, magnetic defects like the monopole, vortex and chain are specified. 

\section{\label{sec:level4}IDENTIFYING MONOPOLES FOR SU($3$) GAUGE GROUP}

One can identify the monopoles of the SU($3$) gauge group directly as well as using its SU($2$) subgroups. The second method is easier since  
the calculations are done for a lower color group and the results have already been reported in our earlier paper \cite{karimi}, as well. We follow the second method 
and study the SU($3$) monopoles. Using the second method, we also study
the relation between SU($2$) subgroups defects and their SU($3$) counterparts.
 
First, we use the appropriate gauge transformations for each of the three SU($2$) subgroups.
Then, we rewrite the results in terms of the local frames and find a connection between this method and the Cho decomposition.
Using the results of the Cho decomposition for SU($3$) group, we study the relation between the magnetic defects obtained in the three SU($2$) 
subgroups with their counterparts in SU($3$) gauge group.
At the end, we write two Lagrangians with the help of both SU($3$) and SU($2$) subgroups and by comparison we learn how 
topological objects of the three SU($2$) subgroups interact with each other. 

We recall that monopoles appear as topological defects which are corresponding to the nontrivial second homotopy group $\Pi_2[SU(3)/U(1)^2]=\mathbb{Z}^2$. 
It means that when SU($3$) group is broken to its Cartan subgroup U$(1)^2$, two types of monopoles appear.

We first discuss about the SU($2$) subgroups. SU($3$) gauge group contains two Abelian directions. Suppose that $\hat{m}_i (i = 1, 2, . . . , 8)$ represents a local 
orthonormal octet basis of SU($3$). The Abelian directions are selected to be $\hat{m}_3=\hat{m}$ and $\hat{m}_8=\hat{m}'$.
The space of SU($3$) group is covered by three SU($2$) subgroups. The corresponding local directions of the internal space can be grouped into the following three 
categories,
\begin{equation}
\begin{split}
&(\hat{m}_1,\hat{m}_2,\hat{m}^1)\   \    \    ,    \    \    \    \hat{m}^1=\hat{m},\\
&(\hat{m}_6,\hat{m}_7,\hat{m}^2)\    \    \    ,    \    \    \    \hat{m}^2=-\frac{1}{2}\hat{m}+\frac{\sqrt{3}}{2}\hat{m},'\\
&(\hat{m}_4,-\hat{m}_5,\hat{m}^3)\   \   \   ,   \    \    \    \hat{m}^3=-\frac{1}{2}\hat{m}-\frac{\sqrt{3}}{2}\hat{m}',
\end{split}
\label{15}
\end{equation}
\\
where $\hat{m}^p(p=1,2,3)$ indicate the Abelian directions of SU($2$) subgroups. As shown in the above equations, they are obtained from the
Abelian directions $\hat{m}$ and $\hat{m}'$ of the SU($3$) gauge group. In order to construct the above three local frames, we need to perform three nontrivial gauge 
transformations.
\\
\paragraph*{\textbf{First subgroup:}}
The generators of the first subgroup are $T_1, T_2$ and $T_3$, where $T_i=\frac{\lambda_i}{2}$ and $\lambda_i$s are Gell-Mann matrices given in Appendix \ref{A}.  
The local color frame $(\hat{m}_1,\hat{m}_2,\hat{m}^1)$ is obtained by applying a nontrivial gauge transformation $U$,
\begin{equation}
U=e^{-i\alpha T_3}e^{-i\beta T_2}e^{-i\gamma T_3} \label{U-trans}.
\end{equation}
And,
\begin{equation}
\begin{split}
UT_1U^{-1}&=\hat{m}_1\cdot\vec{T},\\
UT_2U^{-1}&=\hat{m}_2\cdot\vec{T},\\
UT_3U^{-1}&=\hat{m}^1\cdot\vec{T}.
\end{split}
\end{equation}

The frame $(\hat{m}_1,\hat{m}_2,\hat{m}^1)$ will contain a monopole, if one selects the parameters $\alpha$, $\beta$ and $\gamma$ such that $\alpha=\gamma=\varphi, 
\beta=\theta$, where $\theta$ and $\varphi$ are the polar and azimuthal angles. From Eqn. (\ref{U-trans}) the resulted gauge transformation $U$ is,
\begin{equation}
U=\left(
\begin{array}{ccc}
e^{-i\varphi}\cos\frac{\theta}{2}&-\sin\frac{\theta}{2}&0\\
\sin\frac{\theta}{2}&e^{i\varphi}\cos\frac{\theta}{2}&0\\
0&0&1
\end{array}
\right), \label{U-matrix}
\end{equation}
\newpage
and the gauge field is transformed as the following,
\begin{equation}
\begin{split}
\vec{A}_\mu^U.\vec{T}&=U\left( \vec{A}_\mu.\vec{T}\right) U^{-1}+\frac{i}{g}U\partial_\mu U^{-1}\\
&=U(X_\mu^1 T_1+X_\mu^2 T_2+A_\mu^1T_3)U^{-1}+\frac{i}{g}U\partial_\mu U^{-1}.
\end{split}
\end{equation}
Calculating $\frac{i}{g}U\partial_\mu U^{-1}$ by the explicit form of Eqn. \eqref{U-matrix} and using matrix forms of $T_1, T_2, T_3$, the transformed field is obtained,
\begin{widetext}
\begin{equation}
\begin{split}
\vec{A}_\mu^U\cdot\vec{T}&=X_\mu^1\underbrace{\big((-\sin^2\frac{\theta}{2}+\cos^2\frac{\theta}{2}\cos 2\varphi)T_1+\cos^2\frac{\theta}{2}\sin 2\varphi T_2-\sin\theta\cos\varphi T_3\big)}_{=\hat{m}_1\cdot\vec{T}}\\
&+X_\mu^2\underbrace{\big(-\cos^2\frac{\theta}{2}\sin 2\varphi T_1+(\sin^2\frac{\theta}{2}+\cos^2\frac{\theta}{2}\cos 2\varphi)T_2+\sin\theta \sin\varphi T_3\big)}_{=\hat{m}_2\cdot\vec{T}}\\
&+A_\mu^1\underbrace{\big(\sin\theta \cos\varphi T_1+\sin\theta \sin\varphi T_2+\cos\theta T_3\big)}_{=\hat{m}^1\cdot\vec{T}}\\
&+\underbrace{\frac{1}{g}\big((\sin\varphi\ \partial_\mu\theta-\cos\varphi\sin\theta\ \partial_\mu\varphi)T_1-(\cos\varphi\ \partial_\mu\theta+\sin\varphi\sin\theta\  \partial_\mu\varphi)T_2-(1+\cos\theta)\partial_\mu\varphi T_3\big)}_{=-(C_\mu^{(m^1)}\hat{m}^1+\frac{1}{g}\hat{m}^1\times\partial_\mu\hat{m}^1)\cdot\vec{T}},
\end{split}\label{AU}
\end{equation}
\end{widetext}
where,
\begin{equation}
C_\mu^{(m^1)}=-\frac{1}{g}\hat{m}_1\cdot\partial_\mu\hat{m}_2=\frac{1}{g}(1+\cos\theta)\partial_\mu\varphi. \label{cmu1}
\end{equation}

The first three lines of Eqn. (\ref{AU}) are regular. However, the last line is proportional to $\frac{1}{r}$ which indicates the contribution of the monopole and 
the term $C_\mu^{(m^1)}$ diverges at $\theta=0$, representing a Dirac string. Since we are dealing with static objects, $\partial_\mu\varphi$ is replaced by $\vec{\nabla}\varphi$ 
and the last line of Eqn. (\ref{AU}) can be written as the following,
\begin{equation}
\begin{split}
\frac{1}{g}\frac{1}{r}\big(&(\sin\varphi\ \hat{\theta}-\cos\varphi\ \hat{\varphi})T_1-(\cos\varphi\ \hat{\theta}+\sin\varphi\ \hat{\varphi})T_2\\
&-\frac{1+\cos\theta}{\sin\theta}\hat{\varphi}\ T_3\big).
\label{22}
\end{split}
\end{equation} 
Next, we obtain the magnetic flux corresponding for each term of Eqn. (\ref{22}). These are the fluxes that would penetrate the area inside the 
closed contour $c(r,\theta)\equiv\{(r,\theta,\varphi)|0\leq\varphi<2\pi\}$,
\begin{equation}
\begin{split}
\Phi^{\text{flux}}&=\frac{1}{g}\int_0^{2\pi}r\sin\theta d\varphi\ \hat{\varphi}\cdot\frac{1}{r}(\sin\varphi\ \hat{\theta}-\cos\varphi\ \hat{\varphi})T_1\\
&=-\frac{1}{g}\int_0^{2\pi}\sin\theta \cos\varphi\  d\varphi\  T_1=0,
\end{split}
\end{equation}
\begin{equation}
\begin{split}
\Phi^{\text{flux}}&=-\frac{1}{g}\int_0^{2\pi}r\sin\theta d\varphi\ \hat{\varphi}\cdot\frac{1}{r}(\cos\varphi\ \hat{\theta}+\sin\varphi\ \hat{\varphi})T_2\\
&=-\frac{1}{g}\int_0^{2\pi}\sin\theta \sin\varphi \ d\varphi\ T_2=0,
\end{split}
\end{equation}
\begin{equation}
\begin{split}
&\Phi^{\text{flux}}=-\frac{1}{g}\int_0^{2\pi}r\sin\theta d\varphi\ \hat{\varphi}\cdot\frac{1+\cos\theta}{r\sin\theta}\hat{\varphi}T_3\\
&=-\frac{1}{g}(1+\cos\theta)\int_0^{2\pi}d\varphi\ T_3=-\frac{4\pi}{g}\frac{1+\cos\theta}{2} T_3.
\end{split}\label{flux3}
\end{equation}
From the above equations, it is obvious that the magnetic flux of monopoles has no contribution in the color directions $T_1$ and $T_2$ and the total magnetic flux is 
obtained from the contribution of the monopole in the $T_3$th color direction. 
At $\theta=0$, 
the magnetic flux of a Dirac string that enters a monopole located at the origin $r=0$, is equal to $ -\dfrac{4\pi}{g}T_3 $.

Writing the transformed gauge field from Eqn. (\ref{AU}) in terms of its components in the color space, the following expression for the gauge field is obtained
for the first SU($2$) subgroup, 
\begin{equation}
\vec{A}_\mu ^U=(A_\mu^1- C_\mu^{(m^1)})\hat{m}^1-\frac{1}{g}\hat{m}^1\times\partial_\mu\hat{m}^1+X_\mu ^1 \hat{m}_1+X_\mu^ 2\hat{m}_2. \label{A-Trans}
\end{equation}
Identifying,
\begin{equation}
A_\mu^{(m^1)}=A_\mu ^1- C_\mu^{(m^1)}\     ,   \          \vec{W}_\mu^1= X_\mu^ 1 \hat{m}_1+X_\mu^ 2\hat{m}_2,
\label{x1def}
\end{equation}
we rewrite Eqn. (\ref{A-Trans}):
\begin{equation}
\vec{A}_\mu^U=\underbrace{A_\mu^{(m^1)}\hat{m}^1-\frac{1}{g}\hat{m}^1\times\partial_\mu\hat{m}^1}_{=\hat{A}_\mu^{\left( m^1\right)}}+\vec{W}_\mu^1,
\label{28}
\end{equation}
which is in the form of the extended Cho decomposition written in a frame that contains a monopole and the well-known string singularity.

\begin{figure}[ht]
	\begin{center}
		\centering

		\includegraphics[height=3cm, width=4cm]{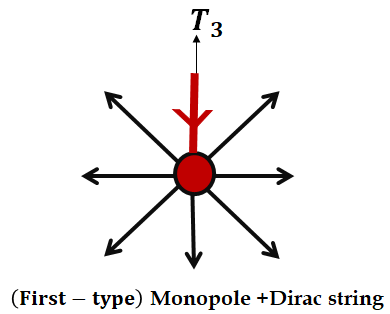}
		\caption{Appearance of a monopole along with a Dirac string for the first subgroup.}
		\label{mon1}

	\end{center}
\end{figure}

The non-Abelian field strength tensor is as the following,
\begin{equation}
	\vec{F}_{\mu\nu}=\partial_\mu \vec{A}_\nu^U-\partial_\nu \vec{A}_\mu^U+g\vec{A}_\mu^U\times \vec{A}_\nu^U.
	\label{29}
\end{equation}
Using $\vec{A}_\mu ^U$ of Eqn. \eqref{28}, the non-Abelian field strength tensor of Eqn. \eqref{29} is,
\begin{equation}
	\vec{F}_{\mu\nu}^1=\underbrace{\left(F_{\mu\nu}^{( m^1)} +H_{\mu\nu}^{( m^1)}\right)\hat{m}^1}_{=\hat{F}_{\mu\nu}^1}+\vec{G}_{\mu\nu}^1+g\vec{W}_\mu^1\times \vec{W}_\nu^1,
	\label{30}
\end{equation}
where,
\begin{equation}\begin{split}
	&F_{\mu\nu}^{( m^1)}=\partial_\mu A_\nu^{( m^1)}-\partial_\nu A_\mu^{( m^1)},\\
	&H_{\mu\nu}^{( m^1)}=\partial_\mu C_\nu^{( m^1)}-\partial_\nu C_\mu^{( m^1)},\\
	&\vec{G}_{\mu\nu}^1=\hat{D}_\mu \vec{W}_\nu^1-\hat{D}_\nu \vec{W}_\mu^1, \quad  \hat{D}_\mu=\partial_\mu+g\hat{A}_\mu^{( m^1)}\times \quad.
	\label{31}
\end{split}\end{equation}
It can be easily confirmed that $H_{\mu\nu}^{( m^1)}$ indicates the field strength of a magnetic monopole sitting at the origin along with a Dirac string 
at $\theta=0$ carrying a magnetic flux equal to $-\frac{4\pi}{g}T_3$. (See Fig. \eqref{mon1}).

We do the same procedure for the two other SU($2$) subgroups to construct the color frames that contain monopoles. 

\paragraph*{\textbf{Second subgroup:}}The generators of the second subgroup 

are defined as $t_1=T_6, t_2=T_7$ and $t_3=-\frac{1}{2}T_3+\frac{\sqrt{3}}{2}T_8$. To construct the color frame $(\hat{m}_6,\hat{m}_7,\hat{m}^2)$, we apply a 
nontrivial gauge transformation $U'$ as the following,
\begin{equation}
U'=e^{-i\alpha t_3}e^{-i\beta t_2}e^{-i\gamma t_3} \label{U-t}.
\end{equation}
And,
\begin{equation}
\begin{split}
U't_1U'^{-1}&=\hat{m}_6\cdot\vec{t},\\
U't_2U'^{-1}&=\hat{m}_7\cdot\vec{t},\\
U't_3U'^{-1}&=\hat{m}^2\cdot\vec{t}.
\end{split}
\end{equation}
In order for the frame $(\hat{m}_6,\hat{m}_7,\hat{m}^2)$ to contain a monopole, we again choose the parameters $\alpha$, $\beta$ and $\gamma$ 
such that $\alpha=\gamma=\varphi, 
\beta=\theta$. Using the $t_i$ of this subgroup in Eqn. \eqref{U-t}, the gauge transformation $U'$ is obtained,
\begin{equation}
U'=\left(
\begin{array}{ccc}
1&0&0\\
0&e^{-i\varphi}\cos\frac{\theta}{2}&-\sin\frac{\theta}{2}\\
0&\sin\frac{\theta}{2}&e^{i\varphi}\cos\frac{\theta}{2}\\
\end{array}
\right). \label{U-matrix2}
\end{equation}
It can be easily shown that the magnetic flux of monopole has no contribution in color
directions $T_6$ and $T_7$ and the total magnetic flux is obtained from the contribution of monopole in $t_3$,
\begin{equation}
\begin{split}
\Phi^{\text{flux}}&=-\frac{1}{g}\int_0^{2\pi}r\sin\theta d\varphi\ \hat{\varphi}\cdot\frac{1+\cos\theta}{r\sin\theta}\hat{\varphi} \ (-\frac{1}{2}T_3+\frac{\sqrt{3}}{2}T_8)\\
&=-\frac{1}{g}(1+\cos\theta)\int_0^{2\pi}d\varphi\ (-\frac{1}{2}T_3+\frac{\sqrt{3}}{2}T_8)\\
&=-\frac{4\pi}{g}\frac{1+\cos\theta}{2} t_3.
\end{split} \label{fluxt3}
\end{equation}
At $\theta=0$, 
the magnetic flux of a Dirac string that enters a monopole located at the origin $r=0$, is equal to $-\dfrac{4\pi}{g}t_3$.

Similar to the calculations of the first subgroup, here the transformed gauge field is,
\begin{equation}
\vec{A}_\mu ^{U'}=(A_\mu^2- C_\mu^{(m^2)})\hat{m}^2-\frac{1}{g}\hat{m}^2\times\partial_\mu\hat{m}^2+X_\mu^ 6 \hat{m}_6+X_\mu^ 7\hat{m}_7,
\end{equation}
where,
\begin{equation}
C_\mu^{(m^2)}=-\frac{1}{g}\hat{m}_6\cdot\partial_\mu\hat{m}_7=\frac{1}{g}(1+\cos\theta)\partial_\mu\varphi. \label{cmu2}
\end{equation}
The components of $\hat{m}_6, \hat{m}_7$ and $\hat{m}^2$ are given in Appendix \ref{compon}. Identifying,
\begin{equation}
A_\mu^{(m^2)}=A_\mu ^2- C_\mu^{(m^2)}\     ,   \          \vec{W}_\mu^2= X_\mu ^6 \hat{m}_6+X_\mu^ 7\hat{m}_7,
\label{x3def}
\end{equation}
we get,
\begin{equation}
\vec{A}_\mu^{U'}=\underbrace{A_\mu^{(m^2)}\hat{m}^2-\frac{1}{g}\hat{m}^2\times\partial_\mu\hat{m}^2}_{=\hat{A}_\mu^{(m^2)}}+\vec{W}_\mu^2,
\label{39a}
\end{equation}
which is again in the form of the extended Cho decomposition written in a frame which contains a monopole and the string singularity.

\begin{figure}[ht]
	\begin{center}
		\centering

		\includegraphics[height=3cm, width=4cm]{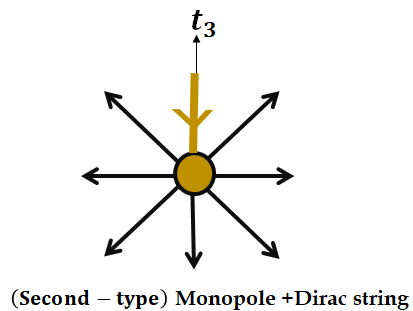}
		\caption{Appearance of a monopole along with a Dirac string for the second subgroup.}
		\label{mon2}

	\end{center}
\end{figure}

Also, similar to the calculations of the first subgroup, the non-Abelian field strength tensor is as the following,
\begin{equation}
	\vec{F}_{\mu\nu}^2=\underbrace{\left(F_{\mu\nu}^{( m^2)} +H_{\mu\nu}^{( m^2)}\right)\hat{m}^2}_{=\hat{F}_{\mu\nu}^2}+\vec{G}_{\mu\nu}^2+g\vec{W}_\mu^2\times \vec{W}_\nu^2,
	\label{40}
\end{equation}
where,
\begin{equation}\begin{split}
		&F_{\mu\nu}^{( m^2)}=\partial_\mu A_\nu^{( m^2)}-\partial_\nu A_\mu^{( m^2)},\\
		&H_{\mu\nu}^{( m^2)}=\partial_\mu C_\nu^{( m^2)}-\partial_\nu C_\mu^{( m^2)},\\
		&\vec{G}_{\mu\nu}^2=\hat{D}_\mu \vec{W}_\nu^2-\hat{D}_\nu \vec{W}_\mu^2, \quad  \hat{D}_\mu=\partial_\mu+g\hat{A}_\mu^{( m^2)}\times \quad,
				\label{41}
\end{split}\end{equation}
$H_{\mu\nu}^{( m^2)}$ indicates the field strength of a magnetic monopole sitting
at the origin along with a Dirac string at $\theta=0$ carrying a magnetic flux equal to $-\frac{4\pi}{g}t_3$. (See Fig. \eqref{mon2}).
\\

\paragraph*{\textbf{Third subgroup:}}The generators of the third SU($2$) subgroup are $t'_1=T_4, t'_2=-T_5$ and $t'_3=-\frac{1}{2}T_3-\frac{\sqrt{3}}{2}T_8$. 
The corresponding nontrivial gauge transformation which leads to a theory that contains the monopole is, 
\begin{equation}
U''=\left(
\begin{array}{ccc}
e^{i\varphi}\cos\frac{\theta}{2}&0&\sin\frac{\theta}{2}\\
0&1&0\\
-\sin\frac{\theta}{2}&0&e^{-i\varphi}\cos\frac{\theta}{2}\\

\end{array}
\right). \label{U-matrix3}
\end{equation}
The total magnetic flux is obtained only from the contribution of the monopole in the color direction $t_3^\prime$,
\begin{equation}
\begin{split}
\Phi^{\text{flux}}&=-\frac{1}{g}\int_0^{2\pi}r\sin\theta d\varphi\ \hat{\varphi}\cdot\frac{1+\cos\theta}{r\sin\theta}\hat{\varphi}(-\frac{1}{2}T_3-\frac{\sqrt{3}}{2}T_8)\\
&=-\frac{1}{g}(1+\cos\theta)\int_0^{2\pi}d\varphi\ (-\frac{1}{2}T_3-\frac{\sqrt{3}}{2}T_8)\\
&=-\frac{4\pi}{g}\frac{1+\cos\theta}{2}t_3^\prime,
\end{split}\label{fluxt'3}
\end{equation}
and the transformed gauge field is,
\begin{equation}
\vec{A}_\mu ^{U''}=(A_\mu^3- C_\mu^{(m^3)})\hat{m}^3-\frac{1}{g}\hat{m}^3\times\partial_\mu\hat{m}^3+X_\mu ^4 \hat{m}_4-X_\mu^ 5\hat{m}_5,
\end{equation}
where,
\begin{equation}
C_\mu^{(m^3)}=-\frac{1}{g}\hat{m}_4\cdot\partial_\mu(-\hat{m}_5)=\frac{1}{g}(1+\cos\theta)\partial_\mu\varphi. \label{cmu3}
\end{equation}
The components of $\hat{m}_4, -\hat{m}_5$ and $\hat{m}^3$ are given in Appendix \ref{compon}. Identifying,
\begin{equation}
A_\mu^{(m^3)}=A_\mu ^3- C_\mu^{(m^3)}\     ,   \          \vec{W}_\mu^3= X_\mu ^4 \hat{m}_4-X_\mu^ 5\hat{m}_5,
\label{x2def}
\end{equation}
we obtain
\begin{equation}
\vec{A}_\mu^{U''}=\underbrace{A_\mu^{(m^3)}\hat{m}^3-\frac{1}{g}\hat{m}^3\times\partial_\mu\hat{m}^3 }_{=\hat{A}_\mu^{\left( m^3\right) }}+\vec{W}_\mu^3,
\end{equation}
which is again in the form of the extended Cho decomposition written in a frame which contains a monopole
and the string singularity.

\begin{figure}[ht]
	\begin{center}
		\centering

		\includegraphics[height=3cm, width=4cm]{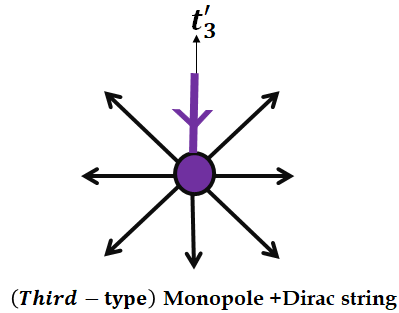}
		\caption{Appearance of a monopole along with a Dirac string for the third subgroup.}
		\label{mon3}

	\end{center}
\end{figure}

The non-Abelian field strength is,
\begin{equation}
	\vec{F}_{\mu\nu}^3=\underbrace{\left(F_{\mu\nu}^{( m^3)} +H_{\mu\nu}^{( m^3)}\right)\hat{m}^3}_{=\hat{F}_{\mu\nu}^3}+\vec{G}_{\mu\nu}^3+g\vec{W}_\mu^3\times \vec{W}_\nu^3,
	\label{48}
\end{equation}
where,
\begin{equation}\begin{split}
		&F_{\mu\nu}^{( m^3)}=\partial_\mu A_\nu^{( m^3)}-\partial_\nu A_\mu^{( m^3)},\\
		&H_{\mu\nu}^{( m^3)}=\partial_\mu C_\nu^{( m^3)}-\partial_\nu C_\mu^{( m^3)},\\
		&\vec{G}_{\mu\nu}^3=\hat{D}_\mu \vec{W}_\nu^3-\hat{D}_\nu \vec{W}_\mu^3, \quad  \hat{D}_\mu=\partial_\mu+g\hat{A}_\mu^{( m^3)}\times \quad,
		\label{49}
\end{split}\end{equation}
$H_{\mu\nu}^{( m^3)}$ indicates the field strength of a magnetic monopole sitting
at the origin along with a Dirac string at $\theta=0$ carrying a magnetic flux equal to $-\frac{4\pi}{g}t'_3$. (See Fig. \eqref{mon3}).

It is clear from Eqns. (\ref{flux3}), (\ref{fluxt3}) and (\ref{fluxt'3}) that there are three magnetic monopoles with the following magnetic charges,
\begin{equation}
	\begin{split}
		g_1&=-\frac{4\pi}{g}T_3,\\
		g_2&=-\frac{4\pi}{g}(-\frac{1}{2}T_3+\frac{\sqrt{3}}{2}T_8)=-\frac{4\pi}{g}t_3,\\
		g_3&=-\frac{4\pi}{g}(-\frac{1}{2}T_3-\frac{\sqrt{3}}{2}T_8)=-\frac{4\pi}{g}t_3^\prime.
	\end{split}\label{gm}
\end{equation}
Given the fact that $g_1+g_2+g_3=0$, only two of the charges are independent. On the other hand, Eqn. (\ref{gm}) can be written in the following form,
\begin{equation}
	g_m=-\frac{4\pi}{g}(w_a\cdot T),
\end{equation}
where $w_a=1,2,3$ are the root vectors of the SU($3$) gauge group and $T$ indicates the vector $(T_3, T_8)$. The magnetic charges are proportional 
to the root vectors and only two of the root vectors are independent. As a result, two of these magnetic charges are independent. The root vectors are 
given in Appendix \ref{A}.

From the above SU($2$) subgroups discussions, the ultimate gauge field describing the monopoles of SU($3$) gauge group can be written as the following,
\begin{equation}
\vec{A}_\mu=\sum_{p=1}^3 \left[ \dfrac{2}{3}(A_\mu^{(m^p)}\hat{m}^p-\frac{1}{g}\hat{m}^p\times\partial_\mu\hat{m}^p)+\vec{W}_\mu^p\right] .
\end{equation}
Therefore,
\begin{equation}\begin{split}
	\vec{A}_\mu&=\dfrac{2}{3}\left\lbrace  A_\mu^{(m^1)}\hat{m}^1+A_\mu^{(m^2)}\hat{m}^2+A_\mu^{(m^3)}\hat{m}^3\right\rbrace  \\
	&-\frac{2}{3}\frac{1}{g}\left\lbrace  \hat{m}^1\times\partial_\mu\hat{m}^1+\hat{m}^2\times\partial_\mu\hat{m}^2+\hat{m}^3\times\partial_\mu\hat{m}^3\right\rbrace  \\
	&+\sum_{p=1}^3\vec{W}_\mu^p.
	\label{53}
\end{split}\end{equation}
Using Eqn. \eqref{15} which shows the relations between SU($3$) Abelian local color frames and its SU($2$) subgroups counterparts, we rewrite the first two 
lines of Eqn. \eqref{53} as follows,
\begin{equation}\begin{split}
		\vec{A}_\mu&= \frac{2}{3} \left( A_\mu^{(m^1)}-\frac{1}{2}A_\mu^{(m^2)}-\frac{1}{2}A_\mu^{(m^3)}\right)\hat{m}\\
		&+\frac{2}{3} \frac{\sqrt{3}}{2} \left( A_\mu^{(m^2)}-A_\mu^{(m^3)}\right)\hat{m}^\prime\\
		&-\frac{1}{g}\left( \hat{m}\times\partial_\mu\hat{m}+\hat{m}^\prime\times\partial_\mu\hat{m}^\prime\right)+\sum_{p=1}^3\vec{W}_\mu^p.
		\label{54}
\end{split}\end{equation}
On the other hand, with the explicit definition of Cho decomposition for SU($3$) gauge group in terms of local color frames $\hat{m}$ and $\hat{m}^\prime$, one may 
write the extended potential $\vec{A}_\mu$ as follows \cite{cho2},
\begin{equation}
		\vec{A}_\mu=\underbrace{A_\mu\hat{m}-\frac{1}{g} \hat{m}\times\partial_\mu\hat{m}}_{=\hat{A}_\mu}+\underbrace{A_\mu^\prime\hat{m}^\prime-\frac{1}{g}\hat{m}^\prime\times\partial_\mu\hat{m}^\prime}_{=\hat{A}_\mu^\prime}+\vec{X}_\mu,
		\label{55}
\end{equation}
where $A_\mu=\hat{m}.\vec{A}_\mu$ and $A_\mu^\prime=\hat{m}^\prime.\vec{A}_\mu^\prime$ are the $T_3$-like and $T_8$-like Abelian components of the 
potential, respectively. $\vec{X}_\mu$ is the gauge covariant part of the potential, called the valence gluon. 

Comparing Eqns. \eqref{54} and \eqref{55}, we find the relation between Abelian components of the potential of the SU($3$) gauge group with
their counterparts for the three SU($2$) subgroups,

\begin{equation}\begin{split}
	&A_\mu=\frac{2}{3} \left( A_\mu^{(m^1)}-\frac{1}{2}A_\mu^{(m^2)}-\frac{1}{2}A_\mu^{(m^3)}\right),\\
	&A_\mu^\prime=\frac{1}{\sqrt{3}}  \left( A_\mu^{(m^2)}-A_\mu^{(m^3)}\right),\\
	&\vec{X}_\mu=\sum_{p=1}^3\vec{W}_\mu^p.
	\label{56}
\end{split}\end{equation}

Also, using the results obtained from the three SU($2$) subgroups, the ultimate field strength describing the SU($3$) gauge group can be written as the following,
\begin{equation}
	\vec{F}_{\mu\nu}=\sum_{p=1}^{3}\left[ \frac{2}{3}\hat{F}_{\mu\nu}^p+\vec{G}_{\mu\nu}^p\right] +g\sum_{p,q}\vec{W}_\mu^p\times \vec{W}_\nu^q.
\end{equation}
Therefore,
\begin{equation}\begin{split}
	\vec{F}_{\mu\nu}&=\frac{2}{3}\left[F_{\mu\nu}^{( m^1) } \hat{m}^1+F_{\mu\nu}^{( m^2) } \hat{m}^2+F_{\mu\nu}^{( m^3) } \hat{m}^3\right] \\
	&+\frac{2}{3}\left[H_{\mu\nu}^{( m^1) } \hat{m}^1+H_{\mu\nu}^{( m^2) } \hat{m}^2+H_{\mu\nu}^{( m^3) } \hat{m}^3\right] \\
	 &+\left[\vec{G}_{\mu\nu}^1+\vec{G}_{\mu\nu}^2+\vec{G}_{\mu\nu}^3 \right] +g\sum_{p,q}\vec{W}_\mu^p\times \vec{W}_\nu^q.
	 \label{58}
\end{split}\end{equation}
Once more, using Eqn. \eqref{15}, 
we rewrite the first two bracket of Eqn. \eqref{58} as the following,
\begin{equation}\begin{split}
		\vec{F}_{\mu\nu}&=\frac{2}{3}\left[F_{\mu\nu}^{( m^1) } -\frac{1}{2}F_{\mu\nu}^{( m^2) }-\frac{1}{2}F_{\mu\nu}^{( m^3) }\right]\hat{m} \\
		&+\frac{1}{\sqrt{3}}\left[ F_{\mu\nu}^{( m^2) }-F_{\mu\nu}^{( m^3) }\right] \hat{m}^\prime\\
		&+\frac{2}{3}\left[H_{\mu\nu}^{( m^1) } -\frac{1}{2}H_{\mu\nu}^{( m^2) }-\frac{1}{2}H_{\mu\nu}^{( m^3) }\right]\hat{m} \\
		&+\frac{1}{\sqrt{3}}\left[ H_{\mu\nu}^{( m^2) }-H_{\mu\nu}^{( m^3) }\right] \hat{m}^\prime\\
		&+\left[\vec{G}_{\mu\nu}^1+\vec{G}_{\mu\nu}^2+\vec{G}_{\mu\nu}^3 \right] +g\sum_{p,q}\vec{W}_\mu^p\times \vec{W}_\nu^q.
		\label{59}
\end{split}\end{equation}
On the other hand, using the explicit definition of the Cho decomposition for SU($3$) gauge group in terms of local frames $\hat{m}$ and $\hat{m}^\prime$,  
the extended filed strength tensor $\vec{F}_{\mu\nu}$ is written as the following,
\begin{equation}
\vec{F}_{\mu\nu}=F_{\mu\nu}\hat{m}+F_{\mu\nu}^\prime\hat{m}^\prime+H_{\mu\nu}\hat{m}+H_{\mu\nu}^\prime\hat{m}^\prime+\vec{K}_{\mu\nu}+g\vec{X}_\mu\times \vec{X}_\nu,
\label{60}
\end{equation}
where,
\begin{equation}\begin{split}
&F_{\mu\nu}=\partial_\mu A_\nu-\partial_\nu A_\mu, \quad F_{\mu\nu}^\prime=\partial_\mu A_\nu^\prime-\partial_\nu A_\mu^\prime\\
&H_{\mu\nu}=\partial_\mu C_\nu-\partial_\nu C_\mu, \quad H_{\mu\nu}^\prime=\partial_\mu C_\nu^\prime-\partial_\nu C_\mu^\prime\\
&\vec{K}_{\mu\nu}=D_\mu \vec{X}_\nu-D_\mu \vec{X}_\nu, \quad D_\mu=\partial_\mu+g\left( \hat{A}_\mu+\hat{A}_\mu^\prime\right) \times\quad.
\label{61}
\end{split}\end{equation}
$C_\mu$ and $C_\mu^\prime$ are the magnetic potentials corresponding to the the magnetic field $H_{\mu\nu}$ and $H_{\mu\nu}^\prime$, directly defined in 
SU($3$) gauge group. 
Comparing Eqns. \eqref{59}, \eqref{60} and \eqref{61}, relations between physical quantities like field strength tensors of SU($3$) gauge group and 
its SU($2$) subgroups are obtained,
\begin{equation}\begin{split}
		&F_{\mu\nu}=\frac{2}{3}\left[F_{\mu\nu}^{( m^1) } -\frac{1}{2}F_{\mu\nu}^{( m^2) }-\frac{1}{2}F_{\mu\nu}^{( m^3) }\right],\\
		&F_{\mu\nu}^\prime=\frac{1}{\sqrt{3}}\left[ F_{\mu\nu}^{( m^2) }-F_{\mu\nu}^{( m^3) }\right] ,\\
		&H_{\mu\nu}=\frac{2}{3}\left[H_{\mu\nu}^{( m^1) } -\frac{1}{2}H_{\mu\nu}^{( m^2) }-\frac{1}{2}H_{\mu\nu}^{( m^3) }\right],\\
		&H_{\mu\nu}^\prime=\frac{1}{\sqrt{3}}\left[ H_{\mu\nu}^{( m^2) }-H_{\mu\nu}^{( m^3) }\right],\\
		&\vec{K}_{\mu\nu}=\vec{G}_{\mu\nu}^1+\vec{G}_{\mu\nu}^2+\vec{G}_{\mu\nu}^3, \\ &g\vec{X}_\mu\times \vec{X}_\nu=g\sum_{p,q}\vec{W}_\mu^p\times \vec{W}_\nu^q.
		\label{62}
\end{split}\end{equation}
Using the fact that $C_{\mu}^{( m^1) } +C_{\mu}^{( m^2) }+C_{\mu}^{( m^3) }=0$, only two of the SU($2$) monopole magnetic potentials are independent. 
Replacing $H_{\mu\nu}$'s of Eqns. \eqref{31}, \eqref{41} and \eqref{49} in $H_{\mu\nu}$ of Eqns. \eqref{62}, we get,
\begin{widetext}\begin{equation}\begin{split}
&H_{\mu\nu}=\frac{2}{3}\left[H_{\mu\nu}^{( m^1) } -\frac{1}{2}H_{\mu\nu}^{( m^2) }-\frac{1}{2}H_{\mu\nu}^{( m^3) }\right]\\
&H_{\mu\nu}=\frac{2}{3}\left[\left( \partial_\mu C_{\nu}^{( m^1) }-\partial_\nu C_{\mu}^{( m^1) }\right)  -\frac{1}{2}\left\lbrace \partial_\mu \left( C_{\nu}^{( m^2) }+ C_{\nu}^{( m^3) }\right) -\partial_\nu \left( C_{\mu}^{( m^2) }+C_{\mu}^{( m^3) }\right) \right\rbrace  \right]
\\&\rightarrow H_{\mu\nu}=H_{\mu\nu}^{(m^1)}\ \text{and} \ H_{\mu\nu}^\prime=\frac{1}{\sqrt{3}}\left[ H_{\mu\nu}^{( m^2) }-H_{\mu\nu}^{( m^3) }\right].
\label{63a}
\end{split}\end{equation}\end{widetext}
The above equations show the relations between the two monopole field strength tensors of the SU($3$) gauge group in terms of their counterparts in the SU($2$) subgroup.

\begin{figure}[ht]
	\begin{center}
		\centering

		\includegraphics[height=3cm, width=4cm]{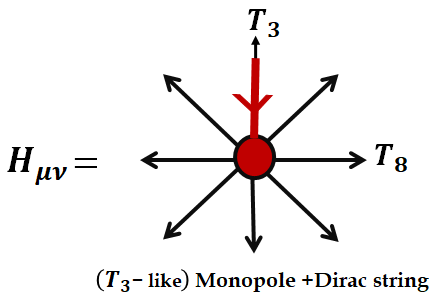}
		\caption{Appearance of  $T_3$-like monopole along with a Dirac string in SU($3$) gauge group.}
		\label{4}

	\end{center}
\end{figure}

Using Figs. \eqref{mon1}, \eqref{mon2}, \eqref{mon3} and  Eqn. \eqref{63a}, we end up to Fig. (\ref{4}).

In fact, figure \eqref{4} shows $H_{\mu\nu}$, one of the monopole field strength tensor of SU($3$). It indicates a magnetic monopole sitting at the origin 
along with a Dirac string in $\theta= 0$ carrying a magnetic flux equal to $-\dfrac{4\pi}{g}T_3$. The interesting point is that it is obtained by SU($2$) 
subgroups and not as a direct calculation of SU($3$) gauge group. 

\begin{figure}[ht]
	\begin{center}
		\centering

		\includegraphics[height=3.5cm, width=8.80cm]{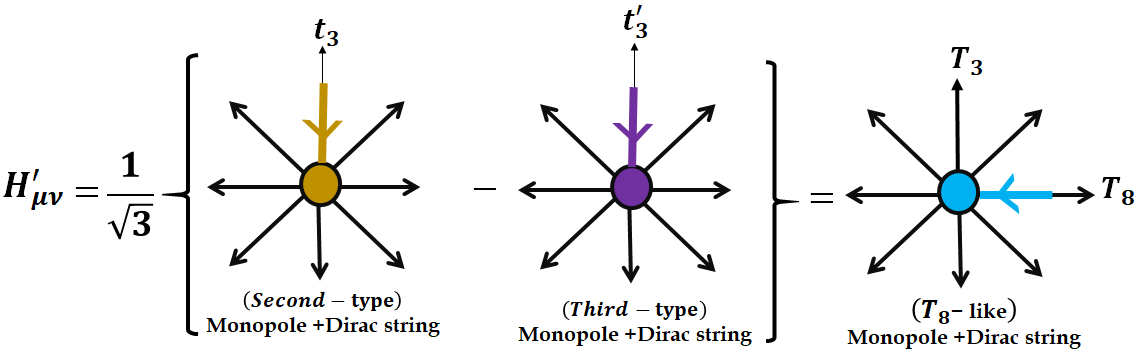}
		\caption{Appearance of  $T_8$-like monopole along with a Dirac string in SU($3$) gauge group.}
		\label{5}

	\end{center}
\end{figure}

In the same way, we get the other monopole field strength tensor of SU($3$), $H_{\mu\nu}^\prime$. It indicates a magnetic monopole sitting at the origin 
along with a Dirac string in $\theta= 0$ carrying a magnetic flux equal $-\dfrac{4\pi}{g}T_8$ (Fig. \eqref{5}).

On the other hand, comparing Eqns. \eqref{31}, \eqref{41}, \eqref{49}, \eqref{61} and \eqref{62}, the relations between the monopole magnetic potential in SU($3$) and its 
three SU($2$) subgroups are as the following,
\begin{equation}\begin{split}
		&C_{\mu}=\frac{2}{3}\left[C_{\mu}^{( m^1) } -\frac{1}{2}C_{\mu}^{( m^2) }-\frac{1}{2}C_{\mu}^{( m^3) }\right],\\
		&C_{\mu}^\prime=\frac{1}{\sqrt{3}}\left[ C_{\mu}^{( m^2) }-C_{\mu}^{( m^3) }\right].
		\label{63}
\end{split}\end{equation}
However, since $C_{\mu}^{( m^1) } +C_{\mu}^{( m^2) }+C_{\mu}^{( m^3) }=0$, only two of the SU($2$) monopole magnetic potentials are independent.
Applying this fact to Eqn. \eqref{63},
\begin{equation}
	C_{\mu}=C_{\mu}^{( m^1) }, \quad
	C_{\mu}^\prime=\frac{1}{\sqrt{3}}\left[ C_{\mu}^{( m^2) }-C_{\mu}^{( m^3) }\right].
	\label{64}
\end{equation}

Therefore, we have found an interesting relation between the defined vector potentials of the two SU($3$) monopoles and their counterparts of the SU($2$) subgroups.

Finally, we discuss about the Lagrangian written directly for the SU($3$) gauge group and compare it with the case where we use its SU($2$) subgroups.
Using field strength tensor of Eqn. \eqref{58}, the extended Lagrangian in terms of the three SU($2$) subgroups is \cite{2019},

\begin{widetext}\begin{equation}\begin{split}
\mathcal{L}&=-\frac{1}{4} \vec{F}_{\mu\nu}.\vec{F}_{\mu\nu}\\
&=\sum_{p=1}^{3}\left\lbrace-\frac{1}{6} (\hat{F}_{\mu\nu}^p)^2-\frac{1}{4}(\vec{G}_{\mu\nu}^p)^2-\frac{g}{2}\hat{F}_{\mu\nu}^p.(\vec{W}_\mu^p \times \vec{W}_\nu^p)\right\rbrace 
-\sum_{p,q} \frac{g^2}{4}(\vec{W}_\mu^p \times \vec{W}_\nu^q)^2-\sum_{p,q,r}\frac{g}{2} \vec{G}_{\mu\nu}^p.(\vec{W}_\mu^q \times \vec{W}_\nu^r)\\
&-\sum_{p\neq q}\frac{g^2}{4}(\vec{W}_\mu^p \times \vec{W}_\nu^q).(\vec{W}_\mu^q \times \vec{W}_\nu^p)
-\sum_{p\neq q}\frac{g^2}{4}(\vec{W}_\mu^p \times \vec{W}_\nu^p).(\vec{W}_\mu^q \times \vec{W}_\nu^q).
\label{65}
\end{split}\end{equation}\end{widetext}
We rewrite the term $-\frac{1}{4}\sum_{p=1}^3(\vec{G}_{\mu\nu}^p)^2$  using $\vec{G}_{\mu\nu}^p$'s
defined in Eqns. (\ref{31}), (\ref{41}) and (\ref{49}) in terms of $\vec{W}_\mu^p$'s; and Eqns. (\ref{x1def}), (\ref{x3def}) and (\ref{x2def}) which 
define $\vec{W}_\mu^p$'s in terms of $ X_\mu$'s.
\begin{widetext}
\begin{equation}
\begin{split}
-\frac{1}{4}\sum_{p=1}^3(\vec{G}_{\mu\nu}^p)^2=
-\frac{1}{4}\bigg\{&\Big[\big(\partial_\mu X_\nu^1-g(A_\mu^{(m^1)}+C_\mu^{(m^1)})X_\nu^2\big)-(\mu\leftrightarrow\nu)\Big]^2+\Big[\big(\partial_\mu X_\nu^2+g(A_\mu^{(m^1)}+C_\mu^{(m^1)})X_\nu^1\big)-(\mu\leftrightarrow\nu)\Big]^2\\
+&\Big[\big(\partial_\mu X_\nu^6-g(A_\mu^{(m^2)}+C_\mu^{(m^2)})X_\nu^7\big)-(\mu\leftrightarrow\nu)\Big]^2+\Big[\big(\partial_\mu X_\nu^7+g(A_\mu^{(m^2)}+C_\mu^{(m^2)})X_\nu^6\big)-(\mu\leftrightarrow\nu)\Big]^2\\
+&\Big[\big(\partial_\mu X_\nu^4-g(A_\mu^{(m^3)}+C_\mu^{(m^3)})X_\nu^5\big)-(\mu\leftrightarrow\nu)\Big]^2+\Big[\big(\partial_\mu X_\nu^5+g(A_\mu^{(m^3)}+C_\mu^{(m^3)})X_\nu^4\big)-(\mu\leftrightarrow\nu)\Big]^2\bigg\}.
\end{split}
\end{equation}
\end{widetext}

Using Eqns. \eqref{62}, \eqref{63} and finding the field strengths of SU($2$) versus the ones for SU($3$), the Lagrangian can be rewritten for the SU($3$) gauge group as follows, 

\begin{widetext}\begin{equation}\begin{split}
	\mathcal{L}&=-\frac{1}{4}F_{\mu\nu}^{\ 2}-\frac{1}{4}F_{\mu\nu}^{\prime\ 2}-\frac{1}{4}H_{\mu\nu}^{\ 2}-\frac{1}{4}H_{\mu\nu}^{\prime\ 2}\\
	&-\frac{1}{2}\left| \left[\partial_\mu+ig(A_\mu+C_\mu) \right] R_\nu-\left[\partial_\nu+ig(A_\nu+C_\nu) \right] R_\mu\right| ^2\\
	&-\frac{1}{2}\left| \left[\partial_\mu-\frac{1}{2}ig(A_\mu+C_\mu)+\frac{\sqrt{3}}{2}ig(A_\mu^\prime+C_\mu^\prime) \right] B_\nu-\left[\partial_\nu-\frac{1}{2}ig(A_\nu+C_\nu)+\frac{\sqrt{3}}{2}ig(A_\nu^\prime+C_\nu^\prime) \right] B_\mu\right| ^2\\
	&-\frac{1}{2}\left| \left[\partial_\mu-\frac{1}{2}ig(A_\mu+C_\mu)-\frac{\sqrt{3}}{2}ig(A_\mu^\prime+C_\mu^\prime) \right] Y_\nu-\left[\partial_\nu-\frac{1}{2}ig(A_\nu+C_\nu)-\frac{\sqrt{3}}{2}ig(A_\nu^\prime+C_\nu^\prime) \right] Y_\mu\right| ^2\\
	&+\text{other\ interactions},
	\label{66}
\end{split}\end{equation}\end{widetext}
where,
\begin{equation}
R_\mu=\frac{X_\mu^1+iX_\mu^2}{\sqrt{2}},\quad B_\mu=\frac{X_\mu^6+iX_\mu^7}{\sqrt{2}},\quad Y_\mu=\frac{X_\mu^4-iX_\mu^5}{\sqrt{2}}.
\end{equation}
Since we are interested in studying the interaction of monopoles, we have just brought the terms that contain the monopole potentials of the second order.
The rest of the Lagrangian is mentioned by \textquotedblleft other interactions\textquotedblright. 
The above Lagrangian is equivalent to the Lagrangian obtained directly by Cho \cite{cho2} who used the field decomposition method for SU($3$) gauge group.
Before proceeding to the interpretation of the Lagrangian components, we should point out the difference between the Cho Lagrangian and \eqref{66}. 
Cho defined the dual magnetic potential $C_\mu^*$ by $H_{\mu\nu}^*=\partial_\mu C_\nu^*-\partial_\nu C_\mu^*$. Where $C_\mu^*$ indicates a timelike potential which has 
no string singularity. However, we need the Dirac string as we explain later. Therefore, we do not remove it at this point. 

The first line of Eqn. \eqref{66} shows the kinetic energy of the Abelian components and the topological defects. The second, the third and the fourth lines show the 
interaction of monopoles with each other. As it is clear from these sentences, each type of monopole can interact with its own type of monopole as well as 
with the other type. These interactions take place via the off-diagonal components of the gauge field. For example, the monopole corresponding to 
the potential $C_\mu$, interacts with the monopole corresponding to the potential $C'_\mu$ via $B_\mu$ and $Y_\mu$. In addition, $C_\mu$ can interact with 
monopoles of its own type via $R_\mu$, $B_\mu$ and $Y_\mu$ and $C'_\mu$ interacts with the monopole of its own type via $B_\mu$ and $Y_\mu$.

\section{\label{sec:level5}IDENTIFYING VORTICES FOR SU($3$)}

As mentioned in section \ref{sec:level2}, to observe the effect of a vortex linking to a Wilson loop, a gauge transformation proportional to the center of the group should be applied.
We use the SU($2$) subgroups of SU($3$) gauge group to identify the SU($3$) vortices. 

\paragraph*{\textbf{First subgroup:}}For this subgroup, the appropriate gauge transformation is,
\begin{equation}
V=e^{-i\xi T_3} \label{vort}.
\end{equation} 
Recalling Eqn. \eqref{S-Gauge} for SU($2$) gauge group, we have,
\begin{equation}
\begin{split}
VT_1V^{-1}&=\hat{n}'_1\cdot\vec{T},\\
VT_2V^{-1}&=\hat{n}'_2\cdot\vec{T},\\
VT_3V^{-1}&=\hat{n}'^1\cdot\vec{T}.
\end{split}
\end{equation}
Choosing $\xi=-\varphi$ in Eqn. (\ref{vort}), the gauge transformation $V$ is,
\begin{equation}
V=\left(
\begin{array}{ccc}
e^{i\frac{\varphi}{2}}&0&0\\
0&e^{-i\frac{\varphi}{2}}&0\\
0&0&1
\end{array}\right) \label{V-T}.
\end{equation}
The transformed gauge field is,
\begin{equation}
\begin{split}
&\vec{A}_\mu^V\cdot\vec{T}=V\vec{A}_\mu\cdot\vec{T}V^{-1}+\frac{i}{g}V\partial_\mu V^{-1}-\text{ideal vortex}\\
&=V(X_\mu^1T_1+X_\mu^2T_2+A_\mu^1T_3)V^{-1}+\frac{i}{g}V\partial_\mu V^{-1}\\&-\text{ideal vortex}.
\end{split}
\end{equation}
As mentioned in section \ref{sec:level2}, the appearance of the last term is due to the discontinuity of gauge 
transformation \cite{engel, rein} and eventually will be removed.
\\

Using $V$ of Eqn. (\ref{V-T}) in the above equation, 
\begin{equation}
\begin{split}
\vec{A}_\mu^V\cdot\vec{T}&=X_\mu^1\underbrace{(\cos\varphi T_1-\sin\varphi T_2)}_{=\hat{n}'_1\cdot \vec{T}}\\
&+X_\mu^2\underbrace{(\sin\varphi T_1+\cos\varphi T_2)}_{=\hat{n}'_2\cdot \vec{T}}+A_\mu^1\underbrace{T_3}_{=\hat{n}'^1\cdot \vec{T}}\\
&+\underbrace{\frac{1}{g}\partial_\mu\varphi T_3}_{=-(C_\mu^{(v^1)}\hat{n}'^1+\frac{1}{g}\hat{n}'^1\times\partial_\mu\hat{n}'^1)},
\end{split}\label{vor1}
\end{equation}
where,
\begin{equation}
C_\mu^{(v^1)}=-\frac{1}{g}\hat{n}'_1\cdot\partial_\mu\hat{n}'_2=-\frac{1}{g}\partial_\mu\varphi,
\end{equation}
\\
and
$\frac{1}{g}\hat{n}'^1\times\partial_\mu\hat{n}'^1=0$. Thus, in terms of the local color frame, the first SU($2$) subgroup gauge field is, 
\begin{equation}
\vec{A}_\mu^V =(A_\mu^1-C_\mu^{(v^1)})\hat{n}'^1+\underbrace{X_\mu^1\hat{n}'_1+X_\mu^2\hat{n}'_2}_{=\vec{W}_\mu^1}.
\end{equation}

The vortex contribution is represented by the last term of Eqn. (\ref{vor1}) and its magnetic flux is obtained as the following,
\begin{equation}
\Phi^{\text{flux}}=\frac{1}{g}\int_0^{2\pi}r\sin\theta d\varphi\ \hat{\varphi}\cdot(\frac{1}{r\sin\theta} \hat{\varphi}\ T_3)=\frac{2\pi}{g}T_3.\label{fv1}
\end{equation} 

\paragraph*{\textbf{Second subgroup:}}For this subgroup, the appropriate gauge transformation is obtained by its generators $t_i$, 
\begin{equation}
V'=\left(
\begin{array}{ccc}
1&0&0\\
0&e^{i\frac{\varphi}{2}}&0\\
0&0&e^{-i\frac{\varphi}{2}}
\end{array}\right).
\end{equation}
And the transformed gauge field is,
\begin{equation}
\vec{A}_\mu^{V'}=(A_\mu^2-C_\mu^{(v^2)})\hat{n}'^2+\underbrace{X_\mu^6\hat{n}'_6+X_\mu^7\hat{n}'_7}_{=\vec{W}_\mu^2},
\end{equation}
where,
\begin{equation}
C_\mu^{(v^2)}=-\frac{1}{g}\hat{n}'_6\cdot\partial_\mu\hat{n}'_7=-\frac{1}{g}\partial_\mu\varphi.
\end{equation}
The components of $\hat{n}'_6, \hat{n}'_7$ and $\hat{n}'^2$ are given in Appendix \ref{compon}. The magnetic flux corresponding to the singular term of the transformed gauge field is,
\begin{equation}
\begin{split}
\Phi^{\text{flux}}&=\frac{1}{g}\int_0^{2\pi}r\sin\theta d\varphi\ \hat{\varphi}\cdot(\frac{1}{r\sin\theta} \hat{\varphi}\ (-\frac{1}{2}T_3+\frac{\sqrt{3}}{2}T_8) )\\
&=\frac{2\pi}{g}(-\frac{1}{2}T_3+\frac{\sqrt{3}}{2}T_8)=\frac{2\pi}{g}t_3.\label{fv2}
\end{split}
\end{equation} 
\\
\paragraph*{\textbf{Third subgroup:}}
Using $t'_i$ for the third SU($2$) subgroup, the appropriate gauge transformation and the gauge field is,
\begin{equation}
V''=\left(
\begin{array}{ccc}
e^{-i\frac{\varphi}{2}}&0&0\\
0&1&0\\
0&0&e^{i\frac{\varphi}{2}}
\end{array}\right).
\end{equation}
And,
\begin{equation}
\vec{A}_\mu^{V''}=(A_\mu^3-C_\mu^{(v^3)})\hat{n}'^3+\underbrace{X_\mu^4\hat{n}'_4-X_\mu^5\hat{n}'_5}_{=\vec{W}_\mu^3},
\end{equation} 
where,
\begin{equation}
C_\mu^{(v^3)}=-\frac{1}{g}\hat{n}'_4\cdot\partial_\mu(-\hat{n}'_5)=-\frac{1}{g}\partial_\mu\varphi.
\end{equation}
The components of $\hat{n}'_4, -\hat{n}'_5$ and $\hat{n}'^3$ are given in Appendix \ref{compon}. The corresponding magnetic flux is,
\begin{equation}
\begin{split}
\Phi^{\text{flux}}&=\frac{1}{g}\int_0^{2\pi}r\sin\theta d\varphi\ \hat{\varphi}\cdot(\frac{1}{r\sin\theta} \hat{\varphi}\ (-\frac{1}{2}T_3-\frac{\sqrt{3}}{2}T_8) )\\
&=\frac{2\pi}{g}(-\frac{1}{2}T_3-\frac{\sqrt{3}}{2}T_8)=\frac{2\pi}{g}t_3^\prime\label{fv3}.
\end{split}
\end{equation} 
From Eqns. (\ref{fv1}), (\ref{fv2}) and (\ref{fv3}), there exist three vortices with the following magnetic fluxes, 
\begin{equation}
\begin{split}
\Phi_1&=\frac{2\pi}{g}T_3,\\
\Phi_2&=\frac{2\pi}{g}(-\frac{1}{2}T_3+\frac{\sqrt{3}}{2}T_8)=\frac{2\pi}{g}t_3,\\
\Phi_3&=\frac{2\pi}{g}(-\frac{1}{2}T_3-\frac{\sqrt{3}}{2}T_8)=\frac{2\pi}{g}t_3^\prime.
\end{split}\label{gv}
\end{equation}
It is clear from Eqn. \eqref{gv} that $\Phi_1+\Phi_2+\Phi_3=0$. Therefore, only two of the vortices are independent. The above equations may be written in 
the compact following form,
\begin{equation}
\Phi_v=\frac{2\pi}{g}(w_a\cdot T),
\end{equation}
where $w_a=1,2,3$ are the root vectors of the SU($3$) gauge group and $T$ is the vector $(T_3, T_8)$. Similar to magnetic charges of monopoles, the 
magnetic fluxes of vortices are proportional to the root vectors and since two of the root vectors are independent, two of these vortices are independent.

We recall that the number of vortices of each SU($N$) group is equal to the number of nontrivial center elements of that group. For example, for the SU($3$) group, there are two nontrivial center elements $z_1=e^{i\frac{2}{3}\pi}$ and $z_2=e^{i\frac{4}{3}\pi}$. This means that we have two kinds of vortices for this group.

We would like to calculate the gauge fields, the corresponding field strength tensors and the Lagrangian density of the SU($3$) vortices. 
From the SU($2$) subgroups, the gauge field which contains the vortices of SU($3$) gauge group can be written as the following,
\begin{equation}
\vec{A}_\mu=\sum_{p=1}^3 \Big[ \dfrac{2}{3}\underbrace{\big((A_\mu^{p}-C_\mu^{(v^p)})\hat{n}'^p\big)}_{=\hat{A}_\mu^p}+\vec{W}_\mu^p\Big] .
\end{equation}
Using the above gauge field, the field strength tensor and Lagrangian density are obtained as follows,
\begin{equation}
\begin{split}
\vec{F}_{\mu\nu}&=\sum_{p=1}^3\dfrac{2}{3}\Big(\partial_\mu(A_\nu^{p}-C_\nu^{(v^p)})-\partial_\nu(A_\mu^{p}-C_\mu^{(v^p)})\Big)\hat{n}'^p\\
&+\sum_{p=1}^3\big(\hat{D}_\mu^p\vec{W}_\nu^p-\hat{D}^p_\nu\vec{W}_\mu^p\big)+g\sum_{p,q}\vec{W}_\mu^p\times\vec{W}_\nu^q,
\end{split}
\label{88}
\end{equation}
where,
\begin{equation}
\hat{D}_\mu^p=\partial_\mu+g\hat{A}_\mu^p\times.
\end{equation}
Using Eqn. \eqref{88}, the Lagrangian in terms of three SU($2$) subgroups is obtained,
\begin{equation}\begin{split}
\mathcal{L}&=-\frac{1}{4} \vec{F}_{\mu\nu}.\vec{F}_{\mu\nu}\\
&=\sum_{p=1}^{3}\left\lbrace-\frac{1}{6} (\hat{F}_{\mu\nu}^p)^2-\frac{1}{4}(\vec{G}_{\mu\nu}^p)^2-\frac{g}{2}\hat{F}_{\mu\nu}^p.(\vec{W}_\mu^p \times \vec{W}_\nu^p)\right\rbrace \\
&-\sum_{p,q} \frac{g^2}{4}(\vec{W}_\mu^p \times \vec{W}_\nu^q)^2-\sum_{p,q,r}\frac{g}{2} \vec{G}_{\mu\nu}^p.(\vec{W}_\mu^q \times \vec{W}_\nu^r)\\
&-\sum_{p\neq q}\frac{g^2}{4}(\vec{W}_\mu^p \times \vec{W}_\nu^q).(\vec{W}_\mu^q \times \vec{W}_\nu^p)\\
&-\sum_{p\neq q}\frac{g^2}{4}(\vec{W}_\mu^p \times \vec{W}_\nu^p).(\vec{W}_\mu^q \times \vec{W}_\nu^q),\label{lagvor}
\end{split}\end{equation}
where,
\begin{equation}
\begin{split}
\hat{F}_{\mu\nu}^p&=\Big(\partial_\mu(A_\nu^{p}-C_\nu^{(v^p)})-\partial_\nu(A_\mu^{p}-C_\mu^{(v^p)})\Big)\hat{n}'^p,\\
\vec{G}_{\mu\nu}^p&=\big(\hat{D}_\mu^p\vec{W}_\nu^p-\hat{D}^p_\nu\vec{W}_\mu^p\big).
\end{split}
\end{equation}

Now, we discuss about the interpretation of each term of the Lagrangian\eqref{lagvor}, 
\begin{itemize}
\item $\sum_{p=1}^{3}-\frac{1}{6} (\hat{F}_{\mu\nu}^p)^2$:

\vspace{0.2cm}
This term contains the kinetic energy of the vortices.

\item $\sum_{p=1}^3-\frac{1}{4}(\vec{G}_{\mu\nu}^p)^2$:

\vspace{0.2cm}
This expression indicates the interaction of the vortices with each other that happens via the off-diagonal components of the gauge field.

\item $\sum_{p=1}^3-\frac{g}{2}\hat{F}_{\mu\nu}^p\cdot(\vec{W}_\mu^p \times \vec{W}_\nu^p)$,\\ 
\item $-\sum_{p,q,r}\frac{g}{2} \vec{G}_{\mu\nu}^p.(\vec{W}_\mu^q \times \vec{W}_\nu^r)$: 

\vspace{0.2cm}
These terms show the interaction of the vortices with the off-diagonal components of the gauge field.

\item $-\sum_{p,q} \frac{g^2}{4}(\vec{W}_\mu^p \times \vec{W}_\nu^q)^2 \times \vec{W}_\nu^p)$, 
\\
\item$-\sum_{p\neq q}\frac{g^2}{4}(\vec{W}_\mu^p \times \vec{W}_\nu^q).(\vec{W}_\mu^q \times \vec{W}_\nu^p)$,\\
\item$-\sum_{p\neq q}\frac{g^2}{4}(\vec{W}_\mu^p \times \vec{W}_\nu^p).(\vec{W}_\mu^q \times \vec{W}_\nu^q)$: 

\vspace{0.2cm}
These terms contain the interaction of the off-diagonal components of the gauge field with each other.
\end{itemize}

\section{\label{sec:level6}CORRELATION OF MONOPOLES AND VORTICES IN SU($3$) GAUGE GROUP}

Applying the two gauge transformations $U$ and $V$ of the previous sections in a successive form, $VU$, we get the configurations that 
include correlated monopoles and vortices. We use the SU($2$) subgroups of SU($3$) gauge group, in the same way as we have done for the monopoles and 
vortices. 
It can be easily shown that the vortex gauge transformation does not change the local color directions that have already been changed by the monopole gauge
transformation, but would just add a phase factor to it.
For example, for the first subgroup, the transformation $V$ has been defined as the following,
\begin{equation}
V=e^{-i\xi \hat{m}^1\cdot \vec{T}}.
\end{equation}
Applying this transformation on a monopole gauge transformation $U$ and by expanding the exponential factor, one gets,
\begin{equation}
\begin{split}
VU&=e^{-i\xi \hat{m}^1\cdot \vec{T}}U=e^{-i\xi UT_3U^{-1}}U\\
&=(1-i\xi UT_3U^{-1}+\frac{(-i\xi)^2}{2!}(UT_3U^{-1})^2+\cdots)U\\
&=U(1-i\xi T_3+\frac{(-i\xi)^2}{2!} (T_3)^2+\cdots)\\
&=U e^{-i\xi T_3}.
\end{split}\label{vu}
\end{equation}

\paragraph*{\textbf{First subgroup:}} we define the color frame of the first subgroup as the following,
\begin{equation}
\begin{split}
(VU)T_1(VU)^{-1}&=\hat{n}_1\cdot\vec{T},\\
(VU)T_2(VU)^{-1}&=\hat{n}_2\cdot\vec{T},\\
(VU)T_3(VU)^{-1}&=\hat{n}^1\cdot\vec{T}.
\end{split}
\end{equation}
Choosing $\xi=-\varphi$ and using \eqref{vu}, the $VU$ transformation matrix and the transformed gauge field are,
\begin{equation}
VU=\left(
\begin{array}{ccc}
\cos\frac{\theta}{2}e^{-i\frac{\varphi}{2}}&-\sin\frac{\theta}{2}e^{-i\frac{\varphi}{2}}&0\\
\sin\frac{\theta}{2}e^{i\frac{\varphi}{2}}&\cos\frac{\theta}{2}e^{i\frac{\varphi}{2}}&0\\
0&0&1
\end{array}\right), \label{UV}
\end{equation}
and,
\begin{equation}
\begin{split}
\vec{A}_\mu^{VU}\cdot\vec{T}&=(VU)\vec{A}_\mu\cdot\vec{T}(VU)^{-1}+\frac{i}{g}(VU)\partial_\mu(VU)^{-1}\\
&=(VU)(X_\mu^1T_1+X_\mu^2T_2+A_\mu^1T_3)(VU)^{-1}\\
&+\frac{i}{g}(VU)\partial_\mu(VU)^{-1}.
\end{split}
\end{equation}
Using the definition of $VU$ from Eqn. (\ref{UV}), the transformed gauge field is given,
\begin{equation}
\begin{split}
\vec{A}_\mu^{VU}\cdot\vec{T}&=X_\mu^1\underbrace{(\cos\theta\cos\varphi T_1+\cos\theta\sin\varphi T_2-\sin\theta T_3)}_{=\hat{n}_1\cdot\vec{T}}\\
&+X_\mu^2\underbrace{(-\sin\varphi T_1+\cos\varphi T_2)}_{=\hat{n}_2\cdot\vec{T}}\\
&+A_{\mu}^1\underbrace{(\sin\theta\cos\varphi T_1+\sin\theta\sin\varphi T_2+\cos\theta T_3)}_{=\hat{n}^1\cdot\vec{T}}\\
&+\underbrace{\frac{1}{g}(\sin\varphi\ \partial_\mu\theta\ T_1-\cos\varphi\ \partial_\mu\theta\ T_2-\partial_\mu\varphi\ T_3)}_{=-\big(C_\mu^{(n^1)}\hat{n}^1+\frac{1}{g}\hat{n}^1\times\partial_\mu\hat{n}^1\big)\cdot\vec{T}},
\end{split}\label{chain1}
\end{equation}
where,
\begin{equation}
\begin{split}
C_\mu^{(n^1)}&=-\frac{1}{g}\hat{n}_1\cdot\partial_\mu\hat{n}_2\\
&=\frac{1}{g}(1+\cos\theta)\partial_\mu\varphi-\frac{1}{g}\partial_\mu\varphi=\frac{1}{g}\cos\theta\partial_\mu\varphi\\
&=C_\mu^{(m^1)}+C_\mu^{(v^1)},																																		
\end{split}
\end{equation}
is the sum of magnetic potentials of a monopole and a vortex. In the following we discuss that the above vector field shows the vector field of a chain. Our
reasoning is based on the flux we would get from this vector field and the contribution of the filed in the Lagrangian which causes the interaction between
the monopoles and vortices. 
Using the above equation in the definition of vector field $\vec{A}_\mu ^{VU}$,  
\begin{equation} 
\vec{A}_\mu ^{VU}=\underbrace{(A_\mu^1- C_\mu^{(n^1)})}_{=A_\mu^{(n^1)}}\hat{n}^1-\frac{1}{g}\hat{n}^1\times\partial_\mu\hat{n}^1+\underbrace{X_\mu^1 \hat{n}_1+X_\mu^2\hat{n}_2}_{=\vec{W}_\mu^1}.\label{avua}
\end{equation}
Given the fact that the magnetic objects of our problem are static, the last line of Eqn. (\ref{chain1}) can be written as follows,
\begin{equation}
\frac{1}{g}\frac{1}{r}(\sin\varphi\ \hat{\theta}\ T_1-\cos\varphi\ \hat{\theta}\ T_2-\frac{1}{\sin\theta}\hat{\varphi}\ T_3),
\end{equation} 
where only the last term has a non-zero contribution to the magnetic flux,
\begin{equation}
\begin{split}
\Phi^{\text{flux}}&=-\int_0^{2\pi}(r\sin\theta\ d\varphi\ \hat{\varphi})\cdot(\frac{1}{g}\frac{1}{r\sin\theta}\hat{\varphi}\ T_3) \\
&=-\frac{2\pi}{g}T_3.
\end{split}
\end{equation}
The above magnetic flux is interpreted as the sum of the magnetic flux of a monopole attached to a Dirac string at $\theta=0$, $(-\frac{4\pi}{g}T_3)$, and the 
magnetic flux of a vortex, $(\frac{2\pi}{g}T_3)$. 
\\

\begin{figure}[ht]
	\begin{center}
		\centering

		\subfloat[]{\includegraphics[height=2.5cm, width=3.5cm]{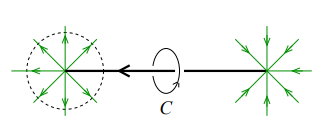}}
		\quad

		\subfloat[]{\includegraphics[height=2.5cm, width=3.5cm]{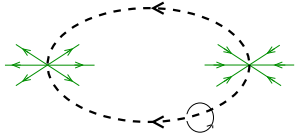}}
		\caption{ Illustration of the connection between the Dirac string shown in plot (a); and the center vortex shown in plot (b) \cite{Rein2001}.
			The interpretation of a chain represented in Fig. \ref{chain11} is the same as this figure where two vortex lines enter a monopole.}
		\label{fig-reind}

	\end{center}
\end{figure}

\begin{figure}[ht]
	\begin{center}
		\centering

		\includegraphics[height=3cm, width=7cm]{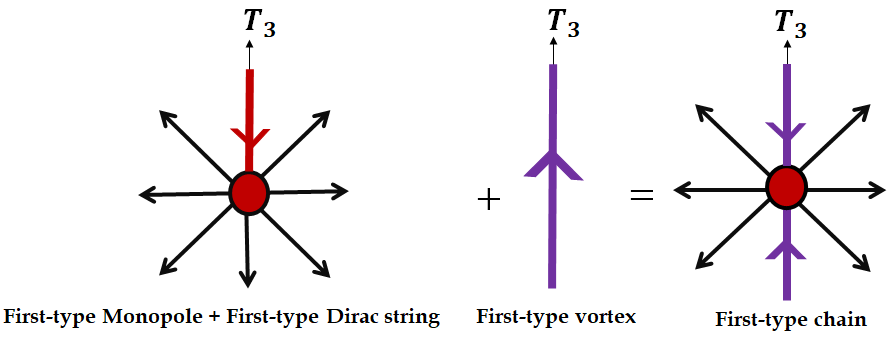}
		\caption{Appearance of a chain of monopole and vortex for the first subgroup.}
		\label{chain11}

	\end{center}
\end{figure}

Using $\vec{A}_\mu ^{VU}$ of Eqn. \eqref{avua}, we get the non-Abelian field strength tensor,
\begin{equation}
	\vec{F}_{\mu\nu}^1=\underbrace{(F_{\mu\nu}^{( n^1)} +H_{\mu\nu}^{(n^1)})\hat{n}^1}_{=\hat{F}_{\mu\nu}^1}+\vec{G}_{\mu\nu}^1+g\vec{W}_\mu^1\times \vec{W}_\nu^1,
	\end{equation}
where,
\begin{equation}\begin{split}
	F_{\mu\nu}^{( n^1)}&=\partial_\mu A_\nu^{( n^1)}-\partial_\nu A_\mu^{( n^1)},\\
	H_{\mu\nu}^{( n^1)}&=\partial_\mu C_\nu^{( n^1)}-\partial_\nu C_\mu^{( n^1)}\\
&=\partial_\mu (C_\nu^{( m^1)}+C_\nu^{(v^1)})-\partial_\nu (C_\mu^{( m^1)}+C_\mu^{(v^1)}),\\
	\vec{G}_{\mu\nu}^1&=\hat{D}_\mu \vec{W}_\nu^1-\hat{D}_\nu \vec{W}_\mu^1, \quad  \hat{D}_\mu=\partial_\mu+g\hat{A}_\mu^{( n^1)}\times \quad.
	\end{split}\end{equation}
It can be easily confirmed that $H_{\mu\nu}^{( n^1)}$ indicates the field strength of a magnetic monopole sitting at the origin along with a Dirac string 
at $\theta=0$ carrying a magnetic flux equal to $-\frac{4\pi}{g}T_3$; plus a line vortex carrying a magnetic flux equal to $\frac{2\pi}{g}T_3$ extending 
on the $z$-axis. The flux carried by the vortex is equal to the half of the flux of a Dirac string, so that the contribution of one Dirac string can be considered to be 
equivalent to the contribution of two vortices. 
At $\theta=0$, half of the Dirac string flux is canceled with the vortex flux located in the positive $z$-axis direction, and a line vortex carrying a magnetic 
flux equal to $\frac{-2\pi}{g}T_3$ remains in the positive $z$-axis. Finally, we have a monopole at the origin, which is connected to two line vortices, a 
line vortex carrying a magnetic flux equal to $\frac{-2\pi}{g}T_3$ on the positive $z$-axis and a line vortex carrying a magnetic flux equal to $\frac{2\pi}{g}T_3$ on the negative $z$-axis, which 
form a chain (See Fig. \eqref{chain11}). 
We have reached to the definition of a chain defined by lattice people who claims that 
the magnetic flux of monopoles at given time is concentrated in a tube-like structure, and the chain is
shorthand terminology for an ordering of monopoles and antimonopoles, in which half the magnetic flux from a monopole ends on
the next antimonopole in the chain, while the other half ends on the previous antimonopole.
In fact, applying the first gauge transformations shown in Eqns. \eqref{U-matrix}, \eqref{U-matrix2} and \eqref{U-matrix3}, we 
have found monopoles for the three subgroups in Eqns. \eqref{cmu1}, \eqref{cmu2} and \eqref{cmu3} that attached 
to their own Dirac strings at $\theta=0$. The appearance of Dirac strings confirms the
existence of the antimonopoles at infinity. This is in agreement with figure $(3)$
of the paper \cite{Rein2001}, which is shown in figure \eqref{fig-reind} in this article. 
Performing the second gauge transformation, the vortex appears along with the monopole which is attached to 
its Dirac sting. In other words, applying two successive gauge transformations like Eqn. \eqref{vu}
gives us all the defects we need. 
 \vspace{0.5cm}

Repeating the same procedure for the second and the third SU($2$) subgroups, the gauge transformations and the gauge fields are obtained as follows. 
\\

\begin{figure}[ht]
	\begin{center}
		\centering

		\includegraphics[height=3cm, width=7cm]{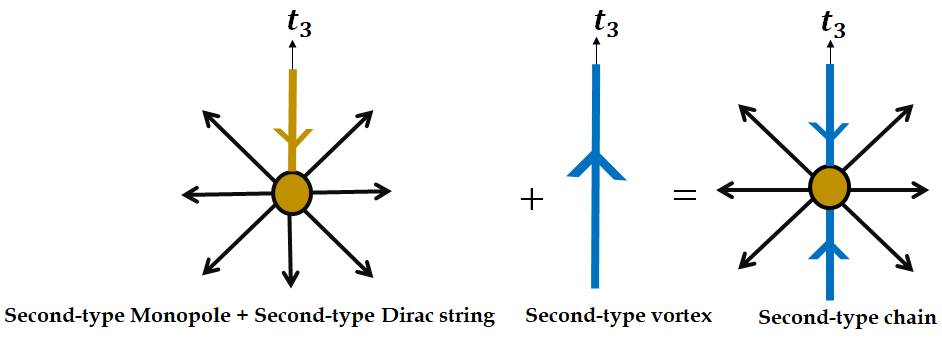}
		\caption{Appearance of a chain of monopole and vortex for the second subgroup.}
		\label{chain22}

	\end{center}
\end{figure}

\begin{figure}[ht]
	\begin{center}
		\centering

		\includegraphics[height=3cm, width=7cm]{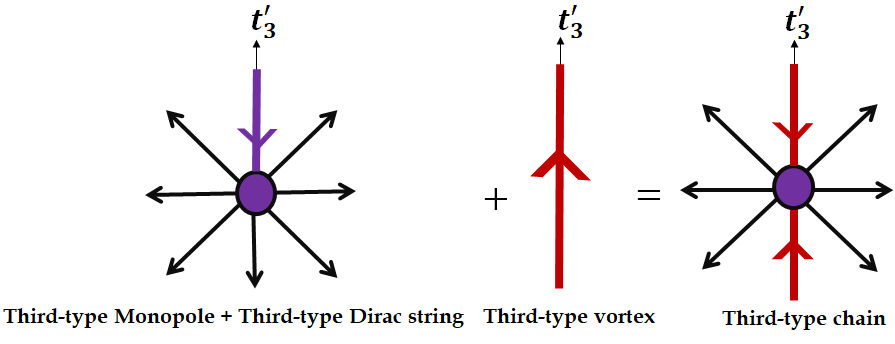}
		\caption{Appearance of a chain of monopole and vortex for the third subgroup.}
		\label{chain33}

	\end{center}
\end{figure}

\paragraph*{\textbf{Second subgroup:}}
For this subgroup:
\begin{equation}
V'U'=\left(
\begin{array}{ccc}
1&0&0\\
0&\cos\frac{\theta}{2}e^{-i\frac{\varphi}{2}}&-\sin\frac{\theta}{2}e^{-i\frac{\varphi}{2}}\\
0&\sin\frac{\theta}{2}e^{i\frac{\varphi}{2}}&\cos\frac{\theta}{2}e^{i\frac{\varphi}{2}}\\
\end{array}\right).
\end{equation}
And,
\begin{equation}
\vec{A}_\mu ^{V'U'}=\underbrace{(A_\mu^2- C_\mu^{(n^2)})}_{=A_\mu^{(n^2)}}\hat{n}^2-\frac{1}{g}\hat{n}^2\times\partial_\mu\hat{n}^2+\underbrace{X_\mu^6 \hat{n}_6+X_\mu^7\hat{n}_7}_{=\vec{W}_\mu^2},\label{avu'}
\end{equation}
where,
\begin{equation}
\begin{split}
C_\mu^{(n^2)}&=-\frac{1}{g}\hat{n}_6\cdot\partial_\mu\hat{n}_7=\frac{1}{g}\cos\theta\partial_\mu\varphi\\
&=C_\mu^{(m^2)}+C_\mu^{(v^2)},
\end{split}
\end{equation}
shows the sum of magnetic potentials of a monopole and a vortex. The components of $\hat{n}_6, \hat{n}_7$ and $\hat{n}^2$ are given in Appendix \ref{compon}. 
Using $\vec{A}_\mu ^{V'U'}$ of Eqn. \eqref{avu'}, we get the non-Abelian field strength tensor,
\begin{equation}
	\vec{F}_{\mu\nu}^2=\underbrace{(F_{\mu\nu}^{( n^2)} +H_{\mu\nu}^{(n^2)})\hat{n}^2}_{=\hat{F}_{\mu\nu}^2}+\vec{G}_{\mu\nu}^2+g\vec{W}_\mu^2\times \vec{W}_\nu^2,
	\end{equation}
where,
\begin{equation}\begin{split}
	F_{\mu\nu}^{( n^2)}&=\partial_\mu A_\nu^{( n^2)}-\partial_\nu A_\mu^{( n^2)},\\
	H_{\mu\nu}^{( n^2)}&=\partial_\mu C_\nu^{( n^2)}-\partial_\nu C_\mu^{( n^2)}\\
&=\partial_\mu (C_\nu^{( m^2)}+C_\nu^{(v^2)})-\partial_\nu (C_\mu^{( m^2)}+C_\mu^{(v^2)}),\\
	\vec{G}_{\mu\nu}^2&=\hat{D}_\mu \vec{W}_\nu^2-\hat{D}_\nu \vec{W}_\mu^2, \quad  \hat{D}_\mu=\partial_\mu+g\hat{A}_\mu^{( n^2)}\times \quad.
	\end{split}\end{equation}
$H_{\mu\nu}^{( n^2)}$ indicates the field strength of a magnetic monopole sitting at the origin along with a Dirac string at $\theta=0$ carrying a magnetic flux 
equal to $-\frac{4\pi}{g}t_3$; plus a line vortex carrying a magnetic flux equal to $\frac{2\pi}{g}t_3$ extending on the $z$-axis. 
This combination makes a chain as explained for the first subgroup. (See Fig. \eqref{chain22}).
\paragraph*{\textbf{Third subgroup:}}
For  this subgroup:
\begin{equation}
V''U''=\left(
\begin{array}{ccc}
\cos\frac{\theta}{2}e^{i\frac{\varphi}{2}}&0&\sin\frac{\theta}{2}e^{i\frac{\varphi}{2}}\\
0&1&0\\
-\sin\frac{\theta}{2}e^{-i\frac{\varphi}{2}}&0&\cos\frac{\theta}{2}e^{-i\frac{\varphi}{2}}\\

\end{array}\right).
\end{equation}
Calculations similar to what is done for the first subgroup leads to the following transformed gauge field,
\begin{equation}
\vec{A}_\mu ^{V''U''}=\underbrace{(A_\mu^3- C_\mu^{(n^3)})}_{A_\mu^{(n^3)}}\hat{n}^3-\frac{1}{g}\hat{n}^3\times\partial_\mu\hat{n}^3+\underbrace{X_\mu ^4 \hat{n}_4-X_\mu^ 5\hat{n}_5}_{\vec{W}_\mu^3},\label{avu''}
\end{equation}
where,
\begin{equation}
\begin{split}
C_\mu^{(n^3)}&=-\frac{1}{g}\hat{n}_4\cdot\partial_\mu(-\hat{n}_5)=\frac{1}{g}\cos\theta\partial_\mu\varphi\\
&=C_\mu^{(m^3)}+C_\mu^{(v^3)},
\end{split}
\end{equation}
indicates the sum of magnetic potentials of a monopole and a vortex. The components of $\hat{n}_4, -\hat{n}_5$ and $\hat{n}^3$ are given in Appendix \ref{compon}. Using $\vec{A}_\mu ^{V''U''}$ of Eqn. \eqref{avu''}, we get the non-Abelian field strength tensor,
\begin{equation}
	\vec{F}_{\mu\nu}^3=\underbrace{(F_{\mu\nu}^{( n^3)} +H_{\mu\nu}^{(n^3)})\hat{n}^3}_{=\hat{F}_{\mu\nu}^3}+\vec{G}_{\mu\nu}^3+g\vec{W}_\mu^3\times \vec{W}_\nu^3,
	\end{equation}
where,
\begin{equation}\begin{split}
	F_{\mu\nu}^{( n^3)}&=\partial_\mu A_\nu^{( n^3)}-\partial_\nu A_\mu^{( n^3)},\\
	H_{\mu\nu}^{( n^3)}&=\partial_\mu C_\nu^{( n^3)}-\partial_\nu C_\mu^{( n^3)}\\
&=\partial_\mu (C_\nu^{( m^3)}+C_\nu^{(v^3)})-\partial_\nu (C_\mu^{( m^3)}+C_\mu^{(v^3)}),\\
	\vec{G}_{\mu\nu}^3&=\hat{D}_\mu \vec{W}_\nu^3-\hat{D}_\nu \vec{W}_\mu^3, \quad  \hat{D}_\mu=\partial_\mu+g\hat{A}_\mu^{( n^3)}\times \quad.
	\end{split}\end{equation}
$H_{\mu\nu}^{( n^3)}$ indicates the field strength of a magnetic monopole sitting at the origin along with a Dirac string at $\theta=0$ carrying a magnetic flux 
equal to $-\frac{4\pi}{g}t'_3$; plus a line vortex carrying a magnetic flux equal to $\frac{2\pi}{g}t'_3$ extending on the $z$-axis, forming a chain (See Fig. \eqref{chain33}).

The corresponding magnetic fluxes for the second and the third subgroups are obtained as follows.
\begin{equation}
\begin{split}
\Phi^{\text{flux}}&=-\frac{1}{g}\int_0^{2\pi}(r\sin\theta\ d\varphi\ \hat{\varphi})\cdot\frac{1}{r\sin\theta}\hat{\varphi}\ t_3\\
&=-\frac{2\pi}{g}t_3.
\end{split}
\end{equation}
The above magnetic flux is interpreted as the sum of the magnetic flux of a monopole attached to a Dirac string at $\theta=0$, $(-\frac{4\pi}{g}t_3)$, 
and the magnetic flux of a vortex, $(\frac{2\pi}{g}t_3)$.
\\

For the third subgroup,
\begin{equation}
\begin{split}
\Phi^{\text{flux}}&=-\frac{1}{g}\int_0^{2\pi}(r\sin\theta\ d\varphi\ \hat{\varphi})\cdot\frac{1}{r\sin\theta}\hat{\varphi}\ t'_3\\
&=-\frac{2\pi}{g}\ t'_3.
\end{split}
\end{equation}
The above magnetic flux is interpreted as the sum of the magnetic flux of a monopole attached to a Dirac string at $\theta=0$, $(-\frac{4\pi}{g}t'_3)$, 
and the magnetic flux of a vortex, $(\frac{2\pi}{g}t'_3)$.

\begin{figure}[h]
	\begin{center}
		\centering

		\subfloat[]{\includegraphics[height=3cm, width=3cm]{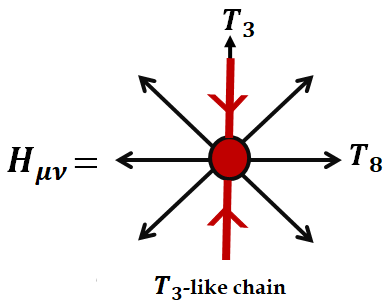}}
		\quad

		\subfloat[]{\includegraphics[height=3cm, width=8.80cm]{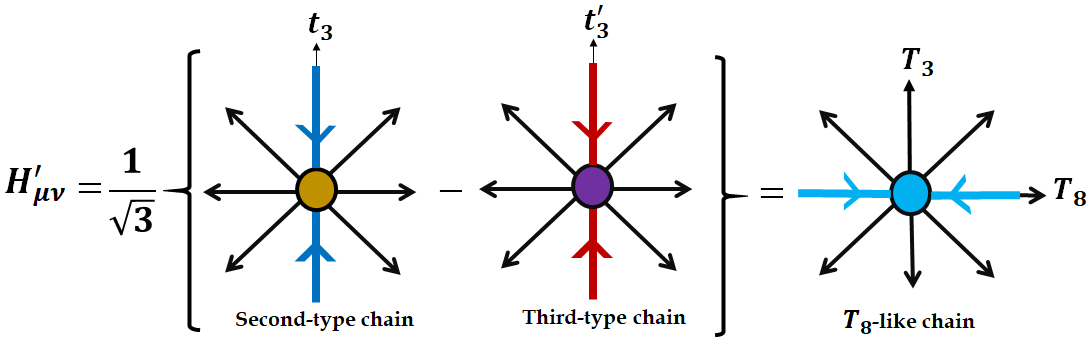}}
		\caption{Appearance of  $T_3$-like chain and $T_8$-like chain in SU($3$) gauge group.}\label{chai}
	\end{center}
\end{figure}

Finally, we reach to the point that we can gather all the information of the SU($2$) subgroups to write a gauge field describing the chains in SU($3$) group, 
\begin{equation}
\begin{split}
\vec{A}_\mu&=\sum_{p=1}^3 \frac{2}{3}(A_\mu^{(n^p)}\hat{n}^p-\frac{1}{g}\hat{n}^p\times\partial_\mu\hat{n}^p)+\vec{W}_\mu^p\\
&=\sum_{p=1}^3 (\frac{2}{3}\hat{A}_\mu^p+\vec{W}_\mu^p)
\end{split}
\end{equation}
where $p$ changes from one to three which represents the three SU($2$) subgroups.
The SU($3$) field strength tensor is obtained from the above gauge filed,
\begin{equation}
\vec{F}_{\mu\nu}=\sum_p(\frac{2}{3}\hat{F}_{\mu\nu}^p+\hat{D}_\mu^p\vec{W}_\nu^p-\hat{D}_\nu^p\vec{W}_\mu^p)+g\sum_{p,q}\vec{W}_\mu^p\times\vec{W}_\nu^q,
\end{equation}
where,
\begin{equation}
\hat{F}_{\mu\nu}^p=\big(\partial_\mu(A_\nu^{(n^p)}+C_\nu^{(n^p)})-\partial_\nu(A_\mu^{(n^p)}+C_\mu^{(n^p)})\big)\hat{n}^p,
\end{equation}
and
\begin{equation}
\hat{D}_\mu^p=\partial_\mu+g\hat{A}_\mu^p\times.
\end{equation}
By performing calculations similar to what was done for monopoles in section \ref{sec:level4}, it is possible to rewrite the magnetic potentials and the field strength tensor 
of the chains in SU($3$) gauge group in terms of their counterparts in SU($2$) subgroups as the following. 
There exist two types of correlations of monopole and vortex, as shown
in figure \eqref{chai}.
\begin{equation}
C_\mu=C_\mu^{(n^1)}\   \   \    , \    \   \  C'_\mu=\frac{1}{\sqrt{3}}(C_\mu^{(n^2)}-C_\mu^{(n^3)}).
\end{equation}
\begin{equation}
H_{\mu\nu}=H_{\mu\nu}^{(n^1)}\   \   \    , \    \   \  H'_{\mu\nu}=\frac{1}{\sqrt{3}}(H_{\mu\nu}^{(n^2)}-H_{\mu\nu}^{(n^3)}).
\end{equation}
\begin{equation}
H_{\mu\nu}=\partial_\mu C_\nu-\partial_\nu C_\mu\   \    ,  \\   \\    H'_{\mu\nu}=\partial_\mu C'_\nu-\partial_\nu C'_\mu.
\end{equation}

Using the above equations, one can argue that there are two types of correlated monopoles and vortices: 
$H_{\mu\nu}$ indicates the field strength of a magnetic monopole sitting at the origin attached to a vortex 
line carrying a flux equal to $-\frac{2\pi}{g}T_3$ at $\theta=0$; and a line vortex carrying a flux equal to $\frac{2\pi}{g}T_3$ at $\theta=\pi$ (See Fig. $\left( \ref{chai}.\text{a}\right).$) 
and $H'_{\mu\nu}$ indicates the field strength of a magnetic monopole sitting at the origin attached to a line vortex carrying a flux equal to $-\frac{2\pi}{g}T_8$ 
at $\theta=0$ and a line vortex carrying a flux equal to $\frac{2\pi}{g}T_8$ at $\theta=\pi$. (Fig. $\left( \ref{chai}.\text{b}\right)$.)

Like monopoles, Lagrangian of chains may be written in two alternative ways: using magnetic potentials of SU($2$) subgroups
$C_\mu^{(n^1)}, C_\mu^{(n^2)}$ and $C_\mu^{(n^3)}$, or SU($3$) magnetic potentials $C_\mu$ and $C'_\mu$.
The chain Lagrangian written with the SU($2$) subgroups is similar to Eqn. \eqref{65}, the Lagrangian with monopole defects.  
The difference is that the $\hat{n}_a$ color frame is used for chains instead of $\hat{m}_a$ which is used for monopole defects; and the vortex 
potential $C_\mu^{(v^a)}$ is added to the monopole potential $C_\mu^{(m^a)}$. 
Since we are interested in studying the interactions between the monopoles and the 
vortices, we only look at those terms of the Lagrangian which are of the order of two with respect to $C_\mu^{(n^a)}$. Those terms include the first and the second 
terms of the Lagrangian of the type \eqref{65} written for chains.
\begin{widetext}
\begin{equation}
\begin{split}
\mathcal{L}=-&\frac{1}{6}\sum_{p=1}^3\Big[\partial_\mu(A_\nu^{(n^p)}+C_\nu^{(m^p)}+C_\nu^{(v^p)})-\partial_\nu(A_\mu^{(n^p)}+C_\mu^{(m^p)}+C_\mu^{(v^p)})\Big]^2\\
-\frac{1}{4}\bigg\{&\Big[\big(\partial_\mu X_\nu^1-g(A_\mu^{(n^1)}+C_\mu^{(m^1)}+C_\mu^{(v^1)})X_\nu^2\big)-(\mu\leftrightarrow\nu)\Big]^2+\Big[\big(\partial_\mu X_\nu^2+g(A_\mu^{(n^1)}+C_\mu^{(m^1)}+C_\mu^{(v^1)})X_\nu^1\big)-(\mu\leftrightarrow\nu)\Big]^2\\
+&\Big[\big(\partial_\mu X_\nu^6-g(A_\mu^{(n^2)}+C_\mu^{(m^2)}+C_\mu^{(v^2)})X_\nu^7\big)-(\mu\leftrightarrow\nu)\Big]^2+\Big[\big(\partial_\mu X_\nu^7+g(A_\mu^{(n^2)}+C_\mu^{(m^2)}+C_\mu^{(v^2)})X_\nu^6\big)-(\mu\leftrightarrow\nu)\Big]^2\\
+&\Big[\big(\partial_\mu X_\nu^4-g(A_\mu^{(n^3)}+C_\mu^{(m^3)}+C_\mu^{(v^3)})X_\nu^5\big)-(\mu\leftrightarrow\nu)\Big]^2+\Big[\big(\partial_\mu X_\nu^5+g(A_\mu^{(n^3)}+C_\mu^{(m^3)}+C_\mu^{(v^3)})X_\nu^4\big)-(\mu\leftrightarrow\nu)\Big]^2\bigg\}\\
+&\text{other\ interactions}
\end{split}
\label{130}
\end{equation}
\end{widetext}
The first term of the above Lagrangian contains the kinetic energy of the monopoles and vortices. It is clear from the Lagrangian that the monopoles of 
each subgroup interact with the monopoles and vortices of the same subgroup via the off-diagonal components of the gauge field. 
If we rewrite the SU($2$) subgroups potentials $C_\mu^{(n^1)}, C_\mu^{(n^2)}$ and $C_\mu^{(n^3)}$ in terms of their counterparts in SU($3$) gauge group: 
$C_\mu$ and $C'_\mu$, and the same for $A_\mu^{(n^1)}, A_\mu^{(n^2)}$ 
and $A_\mu^{(n^3)}$ in terms of $A_\mu$ and $A'_\mu$; we would get the following statements,

\begin{equation}
\begin{split}
C_\mu^{(n^1)}&=C_\mu,\qquad\qquad\qquad\quad\quad A_\mu^{(n^1)}=A_\mu\\
C_\mu^{(n^2)}&=-\frac{1}{2}C_\mu+\frac{\sqrt{3}}{2}C'_\mu, \qquad A_\mu^{(n^2)}=-\frac{1}{2}A_\mu+\frac{\sqrt{3}}{2}A'_\mu,\\
C_\mu^{(n^3)}&=-\frac{1}{2}C_\mu-\frac{\sqrt{3}}{2}C'_\mu, \qquad A_\mu^{(n^3)}=-\frac{1}{2}A_\mu-\frac{\sqrt{3}}{2}A'_\mu.\\
\end{split}
\label{131}
\end{equation}

The last three lines of the Lagrangian Eqn. \eqref{130} contains the interaction between chains.
We will use Eqn. \eqref{131} to study the chains interactions.

The terms including the interaction between chains are as the following,
\begin{itemize}
	\item
	$-\frac{1}{2}g^2C_\mu^{(n^1)}C_\mu^{(n^1)}\left(X_\nu^2X_\nu^2+X_\nu^1X_\nu^1\right)$
	\item
	$+\frac{1}{2}g^2C_\mu^{(n^1)}C_\nu^{(n^1)}\left(X_\mu^2X_\nu^2+X_\mu^1X_\nu^1\right)$
	\item
	$-\frac{1}{2}g^2C_\mu^{(n^2)}C_\mu^{(n^2)}\left(X_\nu^7X_\nu^7+X_\nu^6X_\nu^6\right)$
	\item
	$+\frac{1}{2}g^2C_\mu^{(n^2)}C_\nu^{(n^2)}\left(X_\mu^7X_\nu^7+X_\mu^6X_\nu^6\right)$
	\item
	$-\frac{1}{2}g^2C_\mu^{(n^3)}C_\mu^{(n^3)}\left(X_\nu^5X_\nu^5+X_\nu^4X_\nu^4\right)$
	\item
	$+\frac{1}{2}g^2C_\mu^{(n^3)}C_\nu^{(n^3)}\left(X_\mu^5X_\nu^5+X_\mu^4X_\nu^4\right)$
\end{itemize}
The above equations show that chains of each subgroup interact each other via the the off-diagonal components of their own subgroup. This happens because
the Lagrangian is written for three independent SU($2$) subgroups. However, to get the correct SU($3$) Lagrangian, one should consider that only two of the
SU($2$) subgroup chains are independent and
$C_{\mu}^{( n^1) } +C_{\mu}^{( n^2) }+C_{\mu}^{( n^3) }=0$. Therefore one of them can be written 
in terms of the other two. For example, one can choose $C_\mu^{(n^1)}=-(C_{\mu}^{( n^2) }+C_{\mu}^{( n^3) })$ and then the first two lines of the above 
interactions can be rewritten as follows,
\begin{itemize}
	\item
	$-\frac{1}{2}g^2C_\mu^{(n^2)}C_\mu^{(n^2)}\left(X_\nu^2X_\nu^2+X_\nu^1X_\nu^1\right)$
	\item
	$-\frac{1}{2}g^2C_\mu^{(n^3)}C_\mu^{(n^3)}\left(X_\nu^2X_\nu^2+X_\nu^1X_\nu^1\right)$
	\item
	$-g^2C_\mu^{(n^2)}C_\mu^{(n^3)}\left(X_\nu^2X_\nu^2+X_\nu^1X_\nu^1\right)$
	\item
	$+\frac{1}{2}g^2C_\mu^{(n^2)}C_\nu^{(n^2)}\left(X_\mu^2X_\nu^2+X_\mu^1X_\nu^1\right)$
	\item
	$+\frac{1}{2}g^2C_\mu^{(n^3)}C_\nu^{(n^3)}\left(X_\mu^2X_\nu^2+X_\mu^1X_\nu^1\right)$
	\item
	$+g^2C_\mu^{(n^2)}C_\nu^{(n^3)}\left(X_\mu^2X_\nu^2+X_\mu^1X_\nu^1\right)$
\end{itemize}

The third and sixth lines of the above equations show that the chains of the second and the third subgroups interact  
via the off-diagonal components of the first subgroup, as well as the off-diagonal components of their own subgroup. Therefore, if we choose
$C_\mu^{(n^2)}$ and $C_\mu^{(n^3)}$  magnetic potentials to represent the independent chains, their interactions
can be summarized as the following,
\begin{itemize}
	\item
	$-\frac{1}{2}g^2C_\mu^{(n^2)}C_\mu^{(n^2)}\left(X_\nu^7X_\nu^7+X_\nu^6X_\nu^6+X_\nu^2X_\nu^2+X_\nu^1X_\nu^1\right)$
	\item
	$+\frac{1}{2}g^2C_\mu^{(n^2)}C_\nu^{(n^2)}\left(X_\mu^7X_\nu^7+X_\mu^6X_\nu^6+X_\mu^2X_\nu^2+X_\mu^1X_\nu^1\right)$
	\item
	$-\frac{1}{2}g^2C_\mu^{(n^3)}C_\mu^{(n^3)}\left(X_\nu^5X_\nu^5+X_\nu^4X_\nu^4+X_\nu^2X_\nu^2+X_\nu^1X_\nu^1\right)$
	\item
	$+\frac{1}{2}g^2C_\mu^{(n^3)}C_\nu^{(n^3)}\left(X_\mu^5X_\nu^5+X_\mu^4X_\nu^4+X_\mu^2X_\nu^2+X_\mu^1X_\nu^1\right)$
	\item
	$-g^2C_\mu^{(n^2)}C_\mu^{(n^3)}\left(X_\nu^2X_\nu^2+X_\nu^1X_\nu^1\right)$
	\item
	$+g^2C_\mu^{(n^2)}C_\nu^{(n^3)}\left(X_\mu^2X_\nu^2+X_\mu^1X_\nu^1\right)$
\end{itemize}
The interesting point is that the interaction between the chains of the second and the third subgroups are done via the off-diagonal components of the first subgroup 
while the interaction between two chains of the same subgroup is done via their own off-diagonal components as well as the first subgroup one's.

Using Eqn. (\ref{131}) in Eqn. (\ref{130}), the Lagrangian density in terms of the SU($3$) vector fields is as the following,
 \begin{widetext}\begin{equation}\begin{split}
	\mathcal{L}&=-\frac{1}{4}\Big[\partial_\mu (A_\nu+C_\nu)-\partial_\nu (A_\mu+C_\mu)\Big]^2-\frac{1}{4}\Big[\partial_\mu (A'_\nu+C'_\nu)-\partial_\nu (A'_\mu+C'_\mu)\Big]^2\\
	&-\frac{1}{2}\left| \left[\partial_\mu+ig(A_\mu+C_\mu) \right] R_\nu-\left[\partial_\nu+ig(A_\nu+C_\nu) \right] R_\mu\right| ^2\\
	&-\frac{1}{2}\left| \left[\partial_\mu-\frac{1}{2}ig(A_\mu+C_\mu)+\frac{\sqrt{3}}{2}ig(A_\mu^\prime+C_\mu^\prime) \right] B_\nu-\left[\partial_\nu-\frac{1}{2}ig(A_\nu+C_\nu)+\frac{\sqrt{3}}{2}ig(A_\nu^\prime+C_\nu^\prime) \right] B_\mu\right| ^2\\
	&-\frac{1}{2}\left| \left[\partial_\mu-\frac{1}{2}ig(A_\mu+C_\mu)-\frac{\sqrt{3}}{2}ig(A_\mu^\prime+C_\mu^\prime) \right] Y_\nu-\left[\partial_\nu-\frac{1}{2}ig(A_\nu+C_\nu)-\frac{\sqrt{3}}{2}ig(A_\nu^\prime+C_\nu^\prime) \right] Y_\mu\right| ^2\\
	&+\text{other\ interactions}.
\end{split}\end{equation}\end{widetext}
Where,
\begin{equation}
R_\mu=\frac{X_\mu^1+iX_\mu^2}{\sqrt{2}},\quad B_\mu=\frac{X_\mu^6+iX_\mu^7}{\sqrt{2}},\quad Y_\mu=\frac{X_\mu^4-iX_\mu^5}{\sqrt{2}} 
\end{equation}

The above Lagrangian describes the two different types of monopole and vortex correlations for the SU($3$) gauge group. $C_\mu$ and $C'_\mu$ 
represent two types of chains in SU($3$) gauge group and from the above Lagrangian it is understood that these chains interact with each other via 
the off-diagonal components of the gauge field. We hope to use these information 
to get an effective Lagrangian to study the condensation of vortices or chains for both SU(2) and SU(3) and the confinement. 

\section{\label{sec:level7}CONCLUSIONS}

Many proposals have been suggested to explain confinement which is one of the most challenging problems of QCD. Topological models containing magnetic defects, have 
been able to interpret some of the results given by lattice gauge theories. Monopoles and vortices are among the popular candidates that explain some 
expected characteristic of the quark-antiquark potential. However, none of them have predicted all the features of quark confinement. Chains of monopoles and 
vortices which somehow are observed in lattice calculations, attract people's attention to work more on phenomenological models to compensate the shortcomings of
other candidates. In ref. \cite{karimi}, we discussed about the chains for the SU($2$) gauge group by using two successive gauge transformations proposed in 
\cite{oxman}. As a result of gauge transformations, we defined the local color frames which involve the monopole, the vortex or the chains.
As an extension of our previous work, in this article we have studied the monopoles, vortices and especially chains in SU($3$) using its SU($2$) subgroups. 
Comparison is done with 
Cho decomposition method for SU($3$) monopoles, and in order to get the correct SU($3$) defects, interactions between their SU($2$) counterparts are investigated. 
Using SU($2$) subgroups not only makes the calculations easier but also reveals the possible relations between the SU($3$) defects and their counterparts
in the SU($2$) subgroups. Correlations between monopoles and vortices are discussed for both groups.  One of the task that can be done for the next step
is finding an effective Lagrangian to obtain the confinement potential. It is reasonable to first investigate the condensation of vortices through an effective
Lagrangian which is in progress. 

\appendix
\section{THE SU($3$) GENERATORS AND ROOT VECTORS}\label{A}
The eight Gell-Mann matrices are:
\begin{equation*}
\lambda_1=\left(
\begin{array}{ccc}
0&1&0\\
1&0&0\\
0&0&0
\end{array}\right),  \   \    \lambda_2=\left(\begin{array}{ccc}
0&-i&0\\
i&0&0\\
0&0&0
\end{array}\right), 
\end{equation*}

\begin{equation*}
\lambda_4=\left(\begin{array}{ccc}
0&0&1\\
0&0&0\\
1&0&0
\end{array}\right),  \   \    \lambda_5=\left(\begin{array}{ccc}
0&0&-i\\
0&0&0\\
i&0&0
\end{array}\right), 
\end{equation*}

\begin{equation*}
\lambda_6=\left(
\begin{array}{ccc}
0&0&0\\
0&0&1\\
0&1&0
\end{array}\right),  \   \    \lambda_7=\left(\begin{array}{ccc}
0&0&0\\
0&0&-i\\
0&i&0
\end{array}\right),
\end{equation*}

\begin{equation*}
 \lambda_3=\left(
\begin{array}{ccc}
1&0&0\\
0&-1&0\\
0&0&0
\end{array}\right),\   \    \lambda_8=\frac{1}{\sqrt{3}}\left(
\begin{array}{ccc}
1&0&0\\
0&1&0\\
0&0&-2
\end{array}\right).
\end{equation*}
$T_a=\frac{\lambda_a}{2}$ are the generators of the SU($3$) group. $T_3$ and $T_8$ are two diagonal generators.
\\

The components of root vectors of SU($3$) group are,
\begin{equation*}
w_1=(1,0), \ \ \ w_2=(-\frac{1}{2},-\frac{\sqrt{3}}{2}), \  \   \  w_3=(-\frac{1}{2},\frac{\sqrt{3}}{2}).
\end{equation*}

\section{The components of the local color frames}\label{compon}
The components of the local color frame containing monopoles for the second and the third subgroups are as follows.
\begin{widetext}\begin{equation*}\begin{split}
\hat{m}_6\cdot\vec{t}&=(-\sin^2\frac{\theta}{2}+\cos^2\frac{\theta}{2}\cos 2\varphi)T_6+\cos^2\frac{\theta}{2}\sin 2\varphi T_7-\sin\theta\cos\varphi t_3,\\
\hat{m}_7\cdot\vec{t}&=-\cos^2\frac{\theta}{2}\sin 2\varphi T_6+(\sin^2\frac{\theta}{2}+\cos^2\frac{\theta}{2}\cos 2\varphi)T_7+\sin\theta \sin\varphi t_3,\\
\hat{m}^2\cdot\vec{t}&=\sin\theta \cos\varphi T_6+\sin\theta \sin\varphi T_7+\cos\theta t_3,\\
\hat{m}_4\cdot\vec{t}'&=(-\sin^2\frac{\theta}{2}+\cos^2\frac{\theta}{2}\cos 2\varphi)T_4+\cos^2\frac{\theta}{2}\sin 2\varphi(- T_5)-\sin\theta\cos\varphi t_3^\prime,\\
-\hat{m}_5\cdot\vec{t}'&=-\cos^2\frac{\theta}{2}\sin 2\varphi T_4+(\sin^2\frac{\theta}{2}+\cos^2\frac{\theta}{2}\cos 2\varphi)(-T_5)+\sin\theta \sin\varphi t_3^\prime,\\
\hat{m}^3\cdot\vec{t}'&=\sin\theta \cos\varphi T_4+\sin\theta \sin\varphi (-T_5)+\cos\theta t_3^\prime.\\
\end{split}
\end{equation*}
\end{widetext}
The components of the local color frame containing vortices for the second and the third subgroups are as follows.

\begin{equation*}\begin{split}
\hat{n}'_6\cdot\vec{t}&=\cos\varphi T_6-\sin\varphi T_7,\\
\hat{n}'_7\cdot\vec{t}&=\sin\varphi T_6+\cos\varphi T_7,\\
\hat{n}'^2\cdot\vec{t}&=-\frac{1}{2}T_3+\frac{\sqrt{3}}{2}T_8,\\
\hat{n}'_4\cdot\vec{t}'&=\cos\varphi T_4-\sin\varphi (-T_5),\\
-\hat{n}'_5\cdot\vec{t}'&=\sin\varphi T_4+\cos\varphi (-T_5),\\
\hat{n}'^3\cdot\vec{t}'&=-\frac{1}{2}T_3-\frac{\sqrt{3}}{2}T_8.
\end{split}
\end{equation*}
\\
\\

The components of the local color frame containing correlated monopoles and vortices for the second and the third subgroups are as follows.
\begin{equation*}\begin{split}
\hat{n}_6\cdot\vec{t}&=\cos\theta\cos\varphi T_6+\cos\theta\sin\varphi T_7-\sin\theta t_3,\\
\hat{n}_7\cdot\vec{t}&=-\sin\varphi T_6+\cos\varphi T_7,\\
\hat{n}^2\cdot\vec{t}&=\sin\theta\cos\varphi T_6+\sin\theta\sin\varphi T_7+\cos\theta t_3,\\
\hat{n}_4\cdot\vec{t}'&=\cos\theta\cos\varphi T_4+\cos\theta\sin\varphi (-T_5)-\sin\theta t_3^\prime,\\
-\hat{n}_5\cdot\vec{t}'&=-\sin\varphi T_4+\cos\varphi (-T_5),\\
\hat{n}^3\cdot\vec{t}'&=\sin\theta\cos\varphi T_4+\sin\theta\sin\varphi (-T_5)+\cos\theta t_3^\prime.
\end{split}
\end{equation*}
\\
\\


\newpage
\begin{thebibliography}{99}
\bibitem{pepe}
Ph. de Forcrand and M. Pepe, {\em Nucl. Phys.} {\bf B598} (2001) 557.

\bibitem{green}
J. Greensite, {\em Prog. Part. Nucl. Phys.}{\bf 51} (2003) 1.

\bibitem{chernadub}
M. N. Chernodub and V. I. Zakharov, {\em Phys. Atom. Nucl.} {\bf 72} (2009) 2136.

\bibitem{hosseininejad}
S. M. Hosseini Nejad and S. Deldar, {\em Prog. Theor. Exp. Phys.} {\bf 123} B03 (2016).

\bibitem{hosseininejad2}
S. M. Hosseini Nejad and S. Deldar, {\em Nuclear Physics} {\bf B 917} (2017) 272.

\bibitem{asmaee}
Z. Asmaee, S. Deldar, M. kiamari, {\em Phys. Rev.}{\bf D105} (2022) 096020.

\bibitem{oxman13}
L. E. Oxman, {\em JHEP} 03 (2013) 038.

\bibitem{oxman12}
A. L. L. de Lemos, L. E. Oxman and B. F. I. Teixeira, {\em Phys. Rev.} {\bf D85} (2012) 125014.

\bibitem{oxman19}
L. E. Oxman, {\em Phys. Rev.} {\bf D99} (2019) 016011.

\bibitem{karimi}
N. Karimimanesh, S.Deldar, {\em Int. J. Mod. Phys.} {\bf A37} (2022) 2150255.

\bibitem{duan}
Y.S. Duan and M.L. Ge, {\em Sinica Sci.} {\bf 11} (1979) 1072. 

\bibitem{cho1}
Y. M. Cho, {\em Phys. Rev.} {\bf D21} (1980) 1080.

\bibitem{cho2}
Y. M. Cho, {\em Phys. Rev. Lett.} {\bf 46} (1981) 302.

\bibitem{cho3}
Y. M. Cho, {\em Phys. Rev.} {\bf D23} (1981) 2415.

\bibitem{fadev}
L. Faddeev and A. J. Niemi, {\em Phys. Rev. Lett.}  82 (1999) 1624; {\em Nucl. Phys.} {\bf B776} (2007) 38.

\bibitem{shaban}
S. V. Shabanov, {\em Phys. Lett.} {\bf B458} (1999) 322; {\em Phys. Lett.} {\bf B463} (1999) 263.

\bibitem{massgap}
Y. M. Cho, {\em Int. J. Mod. Phys.} {\bf A29} (2014) 1450013.

\bibitem{2019}
Y. M. Cho,  F. H. Cho, {\em Eur. Phys. J.} {\bf C79} (2019) 498.

\bibitem{Ripka}
G. Ripka, {\em Lect. Notes Phys.} {\bf 639} (2004) 1.

\bibitem{oxman}
L. E. Oxman, {\em JHEP} 12 (2008) 089.






\bibitem{wu}
T. T. Wu and C. N. Yang, {\em Phys. Rev.} {\bf D12} (1975) 3845.

\bibitem{mack}
G. Mack, VB. Petkova, {\em Ann. Phys.} 125 (1980) 117.

\bibitem{engel}
M. Engelhardt, H. Reinhardt, {\em Nucl.Phys.}{\bf B567} (2000) 249. 

\bibitem{rein}
H. Reinhardt, {\em Nucl. Phys.} {\bf B628} (2002) 133.

\bibitem{thinv}
L. E. Oxman, {\em JHEP} 07 (2011) 078.

\bibitem{greensite}
J. Ambjorn, J. Giedt, and J. Greensite, {\em Nucl. Phys. Proc. Suppl.} 83 (2000) 476.‌

\bibitem{zakharov}
F. V. Gubarev, A. V. Kovalenko, M. I. Polikarpov, S. N. Syritsyn, V. I. Zakharov,
{\em Phys. Lett.} {\bf B574} (2003) 136.






\bibitem{Rein2001}
H. Reinhardt and M. Engelhardt, in Quark Confinement and the
Hadron Spectrum IV, W. Lucha and K. M. Maung, eds., pp. 150–162. 
World Scientific, 2002. arXiv:0010031 [hep-th]. 


\end{thebibliography}
\end{document}